\documentclass[12pt]{article}

\usepackage{amsmath,amsfonts,amssymb,cite}

\let\cal\mathcal

\begin{document}
\title {
$${}$$\\
{\bf The Einsteinian T(3)-gauge approach and}\\
{\bf the stress tensor of the screw dislocation }\\
{\bf in the second order:}\\
{\bf avoiding the cut-off at the core} }
\author{
$${}$$\\
{\bf C. MALYSHEV}\\
$${}$$\\
{\it V. A. Steklov Institute of Mathematics,}\\
{\it St.-Petersburg Department,}\\
{\it Fontanka 27, St.-Petersburg, 191023, RUSSIA}\\
{\it E-mail}: malyshev@pdmi.ras.ru\\
$${}$$
}

\date{}

\maketitle

\def \bbe{\boldsymbol\be}
\def \bchi{\boldsymbol\chi}
\def \bdl{\boldsymbol\dl}
\def \bphi{\boldsymbol\phi}
\def \bPsi{\boldsymbol\Psi}
\def \bsi{\boldsymbol\si}
\def \bta{\boldsymbol\eta}
\def \bvphi{\boldsymbol\varphi}
\def \bxi{\boldsymbol\xi}
\def \bna{\boldsymbol\nabla}
\def \binc{\boldsymbol\Inc}

\def \al{\alpha}
\def \be{\beta}
\def \ga{\gamma}
\def \dl{\delta}
\def \ep{\varepsilon}
\def \ze{\zeta}
\def \nb{\nabla}
\def \th{\theta}
\def \ka{\varkappa}
\def \la{\lambda}
\def \si{\sigma}
\def \ph{\varphi}
\def \om{\omega}
\def \Ga{\Gamma}
\def \Dl{\Delta}
\def \La{\Lambda}
\def \Si{\Sigma}
\def \Ph{\Phi}
\def \Om{\Omega}

\def \cA{\cal A}
\def \cB{\cal B}
\def \cC{\cal C}
\def \cD{\cal D}
\def \cE{\cal E}
\def \cN{\cal N}
\def \cI{\cal I}
\def \cT{\cal T}
\def \cR{\cal R}
\def \cF{\cal F}
\def \cY{\cal Y}
\def \cK{\cal K}
\def \cJ{\cal J}
\def \cZ{\cal Z}
\def \cM{{\cal M}}
\def \cL{{\cal L}}
\def \CU{{\cal U}}

\def \BC{\mathbb{C}}
\def \BD{\mathbb{D}}
\def \BZ{\mathbb{Z}}
\def \BR{\mathbb{R}}
\def \BQ{\mathbb{Q}}
\def \IM{\Im}
\def \RE{\Re}
\def \1{^{-1}}
\def \cd{\partial}
\def \at{{\rm arctan}\,}
\def \ch{{\rm ch}\,}
\def \sh{{\rm sh}\,}
\def \th{{\rm th}\,}
\def \bg{{\rm bg}\,}
\def \e{{\rm e}\,}
\def \c{{\rm c}\,}
\def \m{{\rm m}\,}
\def \d{{\rm d}\,}
\def \o{{\rm o}\,}
\def \ld{\ldots}

\def \bQ{{\bf Q}}
\def \vt{\vartheta}
\def \w{\widetilde}
\def \h{\widehat}
\def \t{\times}
\def \r{\rangle}
\def \Tr{{\rm Tr}\,}
\def \Inc{{\rm Inc}\,}
\def \tr{{\rm tr}\,}
\def \diag{{\rm diag}\,}
\def \Det{{\rm det}\,}
\def \z{\zeta}

\def \iso{{\it iso}(3)}
\def \ISO{{\it ISO}(3)}
\def \T{{\it T}(3)}
\def \S{{\it SO}(3)}
\def \ppu{\times\!\!\!\!\!\!\supset}
\def \pps{+\!\!\!\!\!\!\supset}
\def \nt{{\widetilde n}}

\begin{abstract}
\noindent A translational gauge approach of the Einstein type is
proposed for obtaining the stresses that are due to non-singular
screw dislocation. The stress distribution of second order around
the screw dislocation is classically known for the hollow circular
cylinder with traction-free external and internal boundaries. The
inner boundary surrounds the dislocation's core, which is not
captured by the conventional solution. The present gauge approach
enables us to continue the classically known quadratic stresses
inside the core. The gauge equation is chosen in the
Hilbert--Einstein form, and it plays the role of non-conventional
incompatibility law. The stress function method is used, and it
leads to the modified stress potential given by two constituents:
the conventional one, say, the `background' and a short-ranged
gauge contribution. The latter just causes additional stresses,
which are localized. The asymptotic properties of the resulting
stresses are studied. Since the gauge contributions are
short-ranged, the background stress field dominates sufficiently
far from the core. The outer cylinder's boundary is traction-free.
At sufficiently moderate distances, the second order stresses
acquire regular continuation within the core region, and the
cut-off at the core does not occur. Expressions for the
asymptotically far stresses provide self-consistently new length
scales dependent on the elastic parameters. These lengths could
characterize an exteriority of the dislocation core region.
\end{abstract}


\leftline{PACS: 03.50.-z, 11.15.-q, 61.72.Bb, 61.72.Lk}



\thispagestyle{empty}

\newpage

\section{Introduction}

A model of non-singular screw dislocation is proposed in the
present paper. The description is based on the translational gauge
approach of the Einstein type. The aim of the paper is to
investigate the corresponding stress problem in the second order
approximation.

Solutions to the stress problem for the edge and screw
dislocations by means of the theoretical elasticity are known both
in the linear and quadratic approximations (see
\cite{kott1}--\cite{teod}). The stress distributions of the first
order are singular on the dislocation lines. Beyond the scope of
the linear elasticity, the dislocations have been conventionally
considered in \cite{ks}--\cite{wseeg} (the non-linear approaches
are reviewed in \cite{haif} and \cite{gair}). The stress solutions
provided by \cite{ks}--\cite{wseeg} are valid for the hollow
cylinders subject to the traction-free conditions on the external
and internal boundaries. The corresponding internal radii, as the
dislocation core boundaries, remain unspecified. Note that `core'
is rather a crystalline notion since the theoretical elasticity
does not have an appropriate length scale.

The relevance of the second order effects in the theoretical
elasticity to the lattice modelling of the dislocations has been
discussed in \cite{haif}. It is not sufficient to use only the
linear elasticity for the description of strongly distorted
regions of the dislocation cores (see also \cite{hir},
\cite{teod}, \cite{haif}, and \cite{gair}). Self-consistency of
appropriate boundary-value problems at the dislocation cores also
seems theoretically challenging.

The Lagrangian field theory based on the gauge group $\T\ppu\S$
has been proposed in \cite{ed1} and \cite{ed2} as a
non-conventional approach to the defects in the elastic continua.
The Maxwell-type Lagrangian has been proposed to govern the
translational gauge potentials. The ordinary screw dislocation, as
an asymptotic configuration, is allowed for in \cite{ed1} and
\cite{ed2}. Owing to the relevance of higher approximations, it
has been further attempted in \cite{osip} to use \cite{ed1} for
obtaining the corresponding stresses of the second order around
the screw dislocation.

An important modification has later been proposed in \cite{ed3}
for the translational gauge equation advanced in \cite{ed1} and
\cite{ed2}. The modification is concerned with the corresponding
driving source. In its new form, it is given by the difference
between the stress tensor of the model and the stress due to a
classical {\it background} dislocation. In this particular case,
the choice of the background corresponds to the screw dislocation.
The linear approach of \cite{ed3} results in a {\it modified}
screw dislocation, which is equivalent, for sufficiently large
distances, to the ordinary one. It is crucial that additional
short-ranged stresses, which remove the classical stress
singularities, arise. However, it is problematic \cite{mal} to
obtain analogously a non-singular edge dislocation by means of the
translational Lagrangian \cite{ed1} alone.

Further, the translational gauge approach to the elimination of
the dislocation singularities has been developed in
\cite{mal}--\cite{laz4}. The driving source of the gauge equations
in \cite{mal}--\cite{laz4} is the same as in \cite{ed3}, i.e., it
is given in the form of the difference of two stresses. The same
modified screw dislocation as in \cite{ed3} is allowed for in
\cite{mal} and \cite{laz2} as well.

Non-local elasticity should be also recalled as a non-gauge
opportunity for extending the conventional elasticity framework.
Similarly to \cite{ed3}, the singularity-less screw dislocation
has already been independently obtained in \cite{cem} (see also
\cite{cemb}) by means of the non-local elasticity. Moreover, the
first strain gradient elasticity \cite{g1}--\cite{laz5}, as well
as the generalized elasticity \cite{laz6}--\cite{lam2}, also
admits non-singular solutions for the dislocations.

Let us turn again to the gauge approach \cite{osip}, where the
method of stress functions \cite {ks}, \cite{pfl} is used. In
principle, the corresponding Kr\"oner ansatz for the screw
dislocation, in the second order, is the same as for the edge
dislocation, in the first order. However, according to \cite{mal},
the use of the Maxwell-type translational Lagrangian does not
allow for the Airy stress function with a correct numerical
coefficient. Although this Lagrangian is successful for the screw
dislocation in the first order, its usage beyond the scope of the
linear approximation looks doubtful. Moreover, the elastic energy
in \cite{osip} does not contain the terms of third order, which
are accounted for in \cite{ks}, \cite{seegm}, and \cite{pfl}.
There is no continuation of the stresses within the core region.
Therefore, the experience of \cite{osip} looks, to a certain
extent, incomplete.

In principle, the translational gauge approaches discussed in
refs. \cite{mal}--\cite{laz4} look promising for study in the
second order. Indeed, the length scales are ``generated'' in these
models, and the singularities are smoothed out within compact core
regions. The active interest in modelling the defects and their
cores, as well as the importance of the higher approximations,
stimulate us to investigate the gauge models further in order to
gain more realistic descriptions of the core regions.

Specifically, ref. \cite{mal} is based on the Hilbert--Einstein
gauge Lagrangian. Two solutions are found therein, which modify,
for short distances, the Airy and Prandtl dislocational stress
functions. The stress singularities of both classical dislocations
are thus removed. However, certain components of the modified
stresses in the ``edge'' case do not behave properly at infinity.

The approach used in \cite{laz1}--\cite{laz4} is based on the
Lagrangian quadratic in the $\T$-gauge field strength. This
quadratic form is based on three gauge-material
constants\footnote{For a special choice of the parameters, the
gauge Lagrangian in \cite{laz1}--\cite{laz4} is equivalent to the
Lagrangians either in \cite{ed3} or in \cite{mal}. Note that the
most general eight-parameter three-dimensional geometric
Lagrangian is known \cite{kat}, which includes, in addition to the
Hilbert--Einstein term, the terms quadratic in the components of
the differential-geometric torsion and curvature.} and therefore
is different from that in \cite{ed3}. The inclusion of the terms
corresponding to the energy of the rotation gradients enables one
to obtain the modified edge dislocation with an appropriate large
distance behaviour (see \cite{laz4}). The model from \cite{laz4}
contains, in general, ten material constants.

The model \cite{mal} looks attractive for study in the quadratic
approximation. Indeed, the Kr\"oner ansatz for it can be
specified, in each order, appropriately. The gauge approach, say,
\cite{laz1}--\cite{laz4}, would require a rather sophisticated
choice of the gauge parameters for different orders. The
Hilbert--Einsteinian approach implies a single gauge parameter and
looks preferable because of the essential non-linearity of
descriptions based on it. Besides, the corresponding Einstein
tensor is important in the conventional dislocation theory as
well.

In the present paper, the model \cite{mal} is investigated along
the line of \cite{ks}, \cite{seegm}, and \cite{pfl}. The
background defect is given by the straight screw dislocation in
the infinite circular cylinder. Its stress field is taken up to
the second-order terms considered in \cite{pfl}. The Einstein-type
gauge equation plays the role of unconventional incompatibility
law. The stress function approach is used. The stress function of
second order is obtained. It is equal to the sum of the background
and non-conventional parts. The latter results in a continuation
of the stresses within the core region, and the cut-off at the
core's boundary is avoided. The asymptotic properties of the
second order stresses are studied. New length scales of the
``gauge'' origin are provided, which could characterize an
exterior structure of the dislocation core.

We do not discuss in detail a complicated theme such as the
dislocation cores (see \cite{hir}, \cite{teod}, \cite{haif},
\cite{gair}, \cite{cnrs}, and \cite{yam}). For instance,
\cite{pei} and \cite{nab} are the first attempts to incorporate
the lattice periodicity and to display the finiteness of the
cores. Early attempts at treating the screw dislocations and their
cores atomistically are given in \cite{marad}, \cite{doy}, and
\cite{chang}. A singularity-less screw dislocation has also been
obtained in \cite{brail1} by means of the quasi-continuum approach
(see \cite{kun} as well). In their turn, the gauge fields are
important in modern condensed matter physics (see \cite{klei},
\cite{klein}, and \cite{riv1}). Certain applications of \cite{ed1}
and \cite{ed2} can be found in \cite{osip1}--\cite{osip6}.

The paper consists of six sections. Section 1 is introductory (see
also \cite{ed3}--\cite{laz4}). Section 2 outlines the
Einstein-type gauge equation. Section 3 specifies the perturbative
approach. The gauge equation is solved, and the modified stress
potential is obtained in Section 4. The stress fields and their
asymptotics are investigated in Section 5. Section 6 closes the
paper. Details of the calculation are given in Appendices A and B.
The present paper is a refined and improved (concerning the large
distance behaviour of the stresses) version of \cite{ml}.
Bold-faced letters are reserved for second-rank tensors, and
appropriate indices can easily be restored.

\section{The Einstein-type gauge equation  }

The Einsteinian gauge equation is outlined in the present section.
For more on the appropriate differential geometry, one should
refer to \cite{gair}, \cite{klein}, and \cite{lh}. The Eulerian
picture is chosen in the present approach instead of the
Lagrangian one accepted in \cite{mal}. The initial and final
states of the dislocated body are referred to the coordinate
systems $\{x^i\}$ and $\{x^a\}$, respectively. The corresponding
squared length elements are expressed as $g_{i j} dx^idx^j$ and
$\eta_{a b} dx^a dx^b$ (throughout the paper, the indices repeated
imply summation). Let us define the frame components $e^a_{\,\,i}$
by the relation $\cd_{i}=e^a_{\,\,i}\cd_{a}$ (henceforth the
partial derivatives $\cd/\cd x^i$ are denoted by $\cd_i$). The
co-frame components ${\cE}_a^{\,\,i}$ are given by the one-form
$dx^i={\cE}_a^{\,\,i} dx^a$. The components ${\cE}_a^{\,\,i}$ and
their duals $e^a_{\,\,i}$ are orthogonal: $e^a_{\,\,i}
{\cE}_b^{\,\,i}=\dl^a_b$, $\,e^a_{\,\,i} {\cE}_a^{\,\,j}=\dl^j_i$.
For the metric $\eta_{a b}$ we get $\eta_{a b} = g_{i j}
{\cE}_a^{\,\,i}{\cE}_b^{\,\,j}$.

The Eulerian strain tensor is referred to the deformed state, and
it measures a deviation between the final and initial
configurations \cite{gair}. Let us map an initial state to the
deformed one: ${\bf x}\,\longmapsto\,\bxi ({\bf x})$. Then the
difference
$$
\eta_{a b} d\xi^a d\xi^b-g_{i j} dx^idx^j= 2 e_{a b}d\xi^a d\xi^b,
\eqno(2.1)
$$
where
$$
2 e_{a b}\equiv \eta_{a b} - g_{a b},
\qquad g_{a b}\equiv g_{i j} {\cB}_a^{\,\,i}{\cB}_b^{\,\,j}\,,
\eqno(2.2)
$$
defines the Eulerian strain tensor $({\bf e})_{a b}\equiv e_{a
b}$. The metric $g_{a b}$ in (2.2) is called the Cauchy
deformation tensor. Here ${\cB}_a^{\,\,i}$ are given by the
one-form $dx^i={\cB}_a^{\,\,i} d\xi^a$ as follows \cite{gair}:
$$
{\cB}_a^{\,\,i}\,\equiv\,\frac{\cd x^i}{\cd\xi^a}\,=\,
{\cE}_a^{\,\,i}\,-\,{\cE}_b^{\,\,i}\, \stackrel{(\eta)}{\nabla}_a
u^b\,,
\eqno(2.3)
$$
provided that the displacement $\bxi - {\bf x}$ is expanded as
$\xi^i-x^i=u^a {\cE}_a^{\,\,i}$. The covariant derivative
$\stackrel{(\eta)}{\nabla}_a$ in (2.3) is defined by the
requirement that the components ${\cE}_a^{\,\,i}$ are covariantly
constant:
$$
\stackrel{(\eta)}{\nabla}_a\,{\cE}_b^{\,\,i}\,\equiv\, \cd_a
{\cE}_b^{\,\,i}\,-\, \biggl\{\begin{array} {cc} c \\a b\end{array}
                 \biggr\}_\eta\,{\cE}_c^{\,\,i}\,=\,0\,,
\eqno(2.4)
$$
where $\biggl\{\begin{array} {cc} c \\a b\end{array}
\biggr\}_\eta$ is the Christoffel symbol of second kind. The
metric $\eta_{a b}={\cE}_a^{\,\,i}{\cE}_{b\,i}$ is also
covariantly constant because of (2.4). In turn, (2.4) allows one
to express the Christoffel symbols in terms of the metric $\eta_{a
b}$ (see \cite{klein}).

To implement the gauging of the group $T(3)$, we extend the
co-frame components ${\cB}_a^{\,\,i}$ (2.3) by means of the
so-called compensating fields:
$$
{\cB}_a^{\,\,i}\,\equiv\,\frac{\cd x^i}{\cd\xi^a}
\,-\,\ph_a^{\,\,i}\,=\,
{\cE}_a^{\,\,i}\,-\,\Bigl({\cE}_b^{\,\,i}\,
\stackrel{(\eta)}{\nabla}_a u^b\,+\,\ph_a^{\,\,i}\Bigr)\,.
\eqno(2.5)
$$
The entries $\ph_a^{\,\,i}$ are the translational gauge
potentials, which behave under the local transformations
$x^i\,\longrightarrow\,x^i+\eta^i(x)$ as follows:
$$
\begin{array}{lcr}
\displaystyle{
\frac{\cd x^i}{\cd\xi^a}}&\longrightarrow&\displaystyle{
\frac{\cd x^j}{\cd\xi^a}\,\Bigl(\dl^i_j\,+\,\frac{\cd \eta^i}{\cd x^j}
\Bigr)}\,,\\
[0.5cm]
\displaystyle{
\ph_a^{\,\,i}}&\longrightarrow&\displaystyle{\!\!\!\!\!\!\!
\ph_a^{\,\,i}\,+\,\frac{\cd x^j}{\cd\xi^a}\,
\frac{\cd \eta^i}{\cd x^j}}\,.
\end{array}
\eqno(2.6)
$$
The shifts (2.6) ensure the gauge invariance of ${\cB}_a^{\,\,i}$
(2.5) (and of the definitions (2.2), as well). Motivation of (2.5)
should be traced back to $\T\ppu\S$-gauging and to the
corresponding Cartan structure equations \cite{ed1} (see also
\cite{HMMN} and \cite{ml1}). The occurrence of the non-trivial
core region for the screw dislocation in question (i.e., the
appearance of the short-ranged stresses) should be associated with
the gauge variables $\ph_a^{\,\,i}$. The Eulerian strain (2.2)
takes the gauged form:
$$
2 e_{a b}\,=\,\stackrel{(\eta)}{\nabla}_a u_b\,+\,\ph_{a b}
\,+\,\stackrel{(\eta)}{\nabla}_b u_a\,+\,\ph_{b a} \,-\,
\Bigl(\stackrel{(\eta)}{\nabla}_a u_b\,+\,\ph_{a b}\Bigr)
\Bigl(\stackrel{(\eta)}{\nabla}_b u_a\,+\,\ph_{b a}\Bigr) \,.
\eqno(2.7)
$$
When $\bvphi$ is zero, (2.7) is reduced to the
conventionally-looking (background) strain tensor in the Eulerian
picture \cite{gair}.

The differential-geometric approach of the present paper is the
same as in \cite{mal}, i.e., the so-called {\it teleparallel}
framework (see \cite{HMMN} for a quick remind) is adopted as the
most suitable for describing the dislocations \cite{lh}. Let us
consider the Riemann--Christoffel curvature tensor ${{\rm R}_{a b
c}}^d$:
$$
{{\rm R}_{a b c}}^d \,=\, \cd_a \biggl\{\begin{array} {cc} d \\b
c\end{array}
                                        \biggr\}_g\,-\,
\cd_b \biggl\{\begin{array} {cc} d \\a c\end{array}
                                        \biggr\}_g\,+\,
\biggl\{\begin{array} {cc} d \\a e\end{array}
                                         \biggr\}_g
\biggl\{\begin{array} {cc} e \\b c\end{array}
                                         \biggr\}_g\,-\,
\biggl\{\begin{array} {cc} d \\b e\end{array}
                                         \biggr\}_g
\biggl\{\begin{array} {cc} e \\a c\end{array}
                                          \biggr\}_g\,,
\eqno(2.8)
$$
where the Christoffel symbols are calculated for the metric $g_{a
b}$ (2.2) (the subscript `g') given by ${\cB}_a^{\,\,i}$ (2.5).
The metric $g_{a b}$ is covariantly constant, i.e.,
$\stackrel{(g)}{\nabla}_a g_{b c}=0$ is fulfilled, where
$\stackrel{(g)}{\nabla}_a$ is defined similar to (2.4). We obtain
from $\stackrel{(\eta)}{\nabla}_a \eta_{b c}= 0$ the following
relation between the Christoffel symbols:
$$
\displaystyle{ \biggl\{\begin{array} {cc} c \\a b\end{array}
                 \biggr\}_\eta \,-\,
\biggl\{\begin{array} {cc} c \\a b\end{array}
                 \biggr\}_g\,=\,
2 {e_{a b}}^c}\,, \eqno(2.9.1)
$$
$$
2 {e_{a b}}^c\equiv g^{c e}\Bigl( \stackrel{(\eta)}{\nabla}_a e_{b
e} +\stackrel{(\eta)}{\nabla}_b e_{a e}
-\stackrel{(\eta)}{\nabla}_e e_{a b}\Bigr)\,. \eqno(2.9.2)
$$
where (2.2) is taken into account \cite{gair}.

Let us define the Einstein tensor $G^{e
f}\,\equiv\,\displaystyle{\frac 14}\,{\cE}^{e a b} {\cE}^{f c d}
{\rm R}_{a b c d}$, where ${\cE}^{a b c}$ is the Levi--Civita
tensor defined by means of the metric $\eta_{a b}$. Then, the
Einstein-type gauge equation \cite{mal} takes the form
$$
G^{e f}\,=\,\displaystyle{\frac{1}{2 \ell}}\,\bigl(\si^{e f}\,-
\,(\si_{{\rm bg}})^{e f}\bigr)\,. \eqno(2.10)
$$
Here $\ell$ is a constant factor at the Hilbert--Einstein gauge
Lagrangian. The geometry of the deformed state is Euclidean in the
Eulerian approach. The corresponding curvature
${\stackrel{(\eta)}{{\rm R}}_{a b c}}{}^{d}$ is zero for the
metric $\eta_{a b}$. We substitute (2.9.1) into (2.8) and use the
vanishing of ${\stackrel{(\eta)}{{\rm R}}_{a b c}}{}^{d}$. Then,
(2.10) is re-expressed :
$$
-{\cE}^{e a b} {\cE}^{f c d}
\stackrel{(\eta)}{\nabla}_a\,\stackrel{(\eta)}{\nabla}_c \,e_{b
d}\,=\,\displaystyle{\frac{1}{2 \ell}}\,\bigl(\si^{e
f}\,-\,(\si_{{\rm bg}})^{e f}\bigr) \,+\,2 {\cE}^{e a b} {\cE}^{f
c d} e_{a d l} \,{e_{b c}}^{l}\,. \eqno(2.11)
$$
The differential-geometric torsion, as an independent degree of
freedom, is not considered. It is just the Riemann--Christoffel
curvature that is subject to the gauge equation (2.10). Therefore,
(2.11) governs only the variables related to the metric, i.e., the
corresponding strains.

The variational derivation of (2.10) can be discussed along the
line of \cite{mal}. The right-hand side of (2.10) is given by $(2
\ell)^{-1}({\bsi}-{\bsi}_{{\rm bg}})$, where ${\bsi}_{{\rm bg}}$
implies the background stress tensor. The deviation of the stress
tensor of the model from ${\bsi}_{{\rm bg}}$ plays the role of the
driving source in the gauge equation. The parameter $\ell$
characterizes an energy scale intrinsic to the gauge field
${\bvphi}$: it makes the driving source dimensionless. In the
present paper, ${\bsi}_{{\rm bg}}$ implies the stress field of the
straight screw dislocation lying along an infinite cylindric body.
Both ${\bsi}$ and ${\bsi}_{{\rm bg}}$ respect the equilibrium
equations:
$$
\stackrel{(\eta)}{\nabla}_a \si^{a b}=0,\qquad
\stackrel{(\eta)}{\nabla}_a \bigl(\si_{{\rm bg}}\bigr)^{a b}=0\,.
\eqno(2.12)
$$

A more detailed consideration of the gauge geometry behind the
model in question should be done elsewhere, but several refs.
should be mentioned in addition to those listed in \cite{mal}. For
instance, useful indications concerning the translational gauge
geometry can be found in \cite{sard1} and \cite{sard2}. A
topological picture is proposed in \cite{riv}, which includes the
dislocations and extra-matter by means of the torsion and
non-metricity, respectively. Moreover, \cite{mok} provides a
further development of the geometric approach \cite{kat}.

\section{Specification of the gauge equation}
\subsection{The stress function method}

We shall investigate the gauge equation (2.11) using the method of
stress functions proposed in \cite {ks} and \cite{pfl} for solving
the internal stress problems in the incompatible elasticity. An
exposition of \cite{pfl} can be found in ref. \cite{gair} devoted
to a review of the dislocation problems in non-linear elasticity.
Certain details omitted below can be restored with the help of
\cite{pfl} and \cite{gair}.

We shall use the stress function approach in a successive
approximation form. Let us represent the strain and the stress
tensors perturbatively:
$$
{\bf e}\,\approx\,\stackrel{(1)}{{\bf e}}\,+\,\stackrel{(2)}{{\bf
e}}\,, \qquad\qquad {\bsi}\,\approx\,\stackrel{(1)}{{\bsi}}\,+\,
\stackrel{(2)}{{\bsi}}\,, \eqno(3.1)
$$
where $\stackrel{(2)}{{\bf e}}$ and $\stackrel{(2)}{{\bsi}}$ are
of second order smallness in comparison with $\stackrel{(1)}{{\bf
e}}$ and $\stackrel{(1)}{{\bsi}}$. The representation (2.7)
implies that each contribution in (3.1) consists of two parts:
owing to the background and to the non-conventional origin.
Following \cite{pfl}, we shall investigate only the stress
problem. Finding the relationship between the non-conventional
stresses and the gauge potentials is beyond the scope of the
present paper. Substituting (3.1) into (2.11) and (2.12), we
obtain equations of the first ($i=1$) and second ($i=2$) orders:
$$
\displaystyle{
{\bna}{{\boldsymbol\cdot}}\stackrel{(i)}{\bsi}=0}\,, \eqno(3.2.1)
$$
$$
{\binc}\stackrel{(i)}{{\bf e}}\,=\,\displaystyle{\frac{1}{2
\ell}}\,\bdl \stackrel{(i)}{{\bsi}}\,+\,\stackrel{(i-1)}{{\bf
Q}}\,. \eqno(3.2.2)
$$
The tensor notations \cite{gair} are used in (3.2). For instance,
the left-hand side of (3.2.1) takes the form of the divergence of
the tensor of second rank. The notation in (3.2.2) is:
$$
\Bigl({\binc} \stackrel{(i)}{{\bf e}}\Bigr)^{a b} \equiv -{\cal
E}^{a c d} {\cal E}^{b f e} \nabla_c\,\nabla_f\stackrel{(i)}{e}_{d
e}\,, \eqno(3.3.1)
$$
$$
\Bigl(\bdl \stackrel{(i)}{{\bsi}}\Bigr)^{a b}
\equiv\,\stackrel{(i)}{\si}{}^{a b} -
\Bigl(\stackrel{(i)}{{\bsi}}_{{\rm bg}}\Bigr)^{a b}\,,
\eqno(3.3.2)
$$
$$
\stackrel{(0)}{Q}{}^{a b}\equiv 0\,,\qquad \stackrel{(1)}{Q}{}^{a
b}\equiv 2\,{\cal E}^{a c d} {\cal E}^{b f e} \stackrel{(1)}{e}_{c
e l} \stackrel{(1)}{e}_{d  f}{}^l\,, \eqno(3.3.3)
$$
where `${\binc}$' is the double curl {\it incompatibility} {\it
operator}, ${\bna}$ is the covariant derivative
$\stackrel{(\eta)}{\nabla}_a$ $(\equiv\nabla_a)$, and the indices
are raised and lowered by means of the metric ${\bta}$.

Equations (3.2) look similar to the conventional equilibrium and
incompatibility laws: see Eqs. (622) and (623) in \cite{gair}. The
equilibrium equations of the first and second orders are given by
(3.2.1). The gauge equations (3.2.2) play the role of
non-conventional incompatibility conditions. However, there exists
a distinction that is due to $(2 \ell)^{-1} \bdl
\stackrel{(i)}{{\bsi}}$ in (3.2.2). This term is just responsible
for the short-ranged behaviour of the gauge parts of the resulting
stress functions. Moreover, $\stackrel{(1)}{\bf Q}$ in (3.3.3) is
free from the torsion tensor in our approach.

The elastic energy of an isotropic body is chosen in the Eulerian
representation as follows:
$$
W({\bf e})\,=\,j I^2_1 ({\bf e})\,+\, k I_2 ({\bf
e})\,+\,l^{\prime} I^3_1 ({\bf e}) \,+\,m^\prime I_1 ({\bf
e})\,I_2 ({\bf e})\,+ \,n^{\prime} I_3 ({\bf e})\,, \eqno(3.4)
$$
where $j=\mu +\la/2$ and $k=-2\mu$ are the elastic moduli of
second order ($\la$ and $\mu$ are the Lam\'e constants), and
$l^{\prime}$, $m^{\prime}$, and $n^{\prime}$ are the elastic
moduli of third order. Once $W({\bf e})$ has been chosen in the
form (3.4), the constitutive law relates $\stackrel{(i)}{\bf e}$
with $\stackrel{(i)}{\bsi}$ ($i=1, 2$) as follows \cite{pfl}:
$$
\stackrel{(i)}{\bf e}\,=\, C_1 I_1\bigl(\stackrel{(i)}{\bsi}\bigr)
{\bta}\,+\,C_4 \stackrel{(i)}{\bsi} \,+\,\stackrel{(i-1)}{\bPsi}
\,, \eqno(3.5)
$$
where $\stackrel{(0)}{\bPsi}\equiv 0$ and
$$
\stackrel{(1)}{\bPsi}\,\equiv \Bigl( C_2
I^2_1\bigl(\stackrel{(1)}{\bsi}\bigr)\,+\, C_3
I_2\bigl(\stackrel{(1)}{\bsi}\bigr)\Bigr){\bta} \,+\,C_5
I_1\bigl(\stackrel{(1)}{\bsi}\bigr) \stackrel{(1)}{\bsi}\,+ \,C_7
I_3\bigl(\stackrel{(1)}{\bsi}\bigr)
\Bigl(\stackrel{(1)}{\bsi}\Bigr)^{-1}\,.\eqno(3.6)
$$
Here, the $I_{1, 2, 3} (\cdot)$ are the tensor invariants
\cite{mur}. The numerical coefficients $C_1$ and $C_4$ are:
$$
\displaystyle{ C_1\,=\,- (2\mu)^{\1}\frac{\nu}{1+\nu}\,, \qquad
C_4\,=\,(2\mu)^{\1}}\,, \eqno(3.7)
$$
where $\nu=\la/(2(\la+\mu))$ is the Poisson ratio. The
coefficients $C_2$, $C_3$, $C_5$, and $C_7$ can be expressed (see
\cite{pfl} and \cite{gair}) in terms of the elastic moduli of
second and third order (3.4), although this is not used below.
More details on the relation of $\la$, $\mu$ and $l^\prime$,
$m^\prime$, $n^\prime$ with the elastic moduli of crystals can be
found in \cite{teod} and \cite{gair}.

The equilibrium equations (3.2.1) should be fulfilled with the
help of the Kr\"oner ansatz:
$$
\stackrel{(i)}{{\bsi}}\,=\,{\binc} \stackrel{(i)}{{\bchi}}.
\eqno(3.8)
$$
Substituting the constitutive relations (3.5) into (3.2.2) and
using (3.8), we obtain equations for the components of the stress
potential $\stackrel{(i)}{{\bchi}}$:
$$
\begin{array}{c}
\displaystyle{
\Dl \Dl \stackrel{(i)}{{\chi}}_{a b}\,+\,
a\Bigl(\nb_a\nb_b\,-\,\eta_{a b}\Dl\Bigr)
 \Dl I_1\bigl(\stackrel{(i)}{{\bchi}}\bigr)\,+\,
\Bigl(\bigl(1-a\bigr)\nb_a\nb_b\,+\,a\,\eta_{a b}
\Dl \Bigr)\nb^c\nb^d \stackrel{(i)}{{\chi}}_{c d}}\,-\\[0.5cm]
-\,\Dl \Bigl(\nb_a\nb_c \stackrel{(i)}{{\chi}}{}^c_{\,\,\,b}
\,+\,\nb_b\nb_c \stackrel{(i)}{{\chi}}{}^c_{\,\,\,a}\Bigr)
\,=\,\kappa^2 \Bigl(\bdl \stackrel{(i)}{{\bsi}}\Bigr)_{a b}\,
+\,2\mu\stackrel{(i)}{S}{}_{(a b)}\,,
\end{array}
\eqno(3.9)
$$
where
$$
\stackrel{(1)}{S}{}_{a b}\,\equiv\,0\,,\qquad
\stackrel{(2)}{S}{}_{(a b)}\,\equiv\, \stackrel{(1)}{Q}{}_{(a
b)}\,-\, \Bigl({\binc}\stackrel{(1)}{\bPsi}\Bigr)_{(a b)}\,,
$$
$\Delta$ is the Laplacian, and the $\bdl \stackrel{(i)}{{\bsi}}$
are expressed by means of (3.3.2) and (3.8). In addition,
$\stackrel{(1)}{\bf Q}$ and $\stackrel{(1)}{\bPsi}$ are given by
(3.3.3) and (3.6), accordingly. The curly brackets around the
indices imply symmetrization, and we use the dimensionless
parameters $\kappa^2 \equiv \mu/\ell$ and $a$ (see \cite{ed1} and
\cite{ed2}):
$$
a\equiv\frac{\,\la\,}{3\la+2\mu}
\,=\,\frac{\,1\,}{1+{\nu}^{-1}}\,,\qquad
1-a=\frac{\,2(\la+\mu)\,}{3\la+2\mu} \,=\,\frac{\,1\,}{1+\nu}\,.
\eqno(3.10)
$$

\subsection{The gauge equations in the first and second orders}

Let us adjust (3.9) to the special problem in question. We replace
the derivatives ${\nb}_a$ by the partial derivatives $\cd_a$,
where $x^a$ are the coordinates in the final state, and assume
$\cd_3\equiv 0$ (see \cite{pfl} and \cite{gair}). Conventional
notation is adopted for the components of the stress potential
which are non-trivial \cite{lh}:
$$
\begin{array}{l}
\displaystyle{
\mu\stackrel{(i)}{\phi}\,\equiv\,
\cd_2 \stackrel{(i)}{{\chi}}_{1 3}\,-\,
      \cd_1 \stackrel{(i)}{{\chi}}_{2 3}}\,,\qquad\qquad
      i=1,\,2\,,\\ [0.4cm]
f\,\equiv\,\stackrel{(2)}{{\chi}}_{3 3}\,,\quad
p\,\equiv\,-\cd^2_{1 1} \stackrel{(2)}{{\chi}}_{2 2}\,-\, \cd^2_{2
2} \stackrel{(2)}{{\chi}}_{1 1}\,+\, 2\,\cd^2_{1 2}
\stackrel{(2)}{{\chi}}_{1 2}\,.
\end{array}
\eqno(3.11)
$$
The other components of $\stackrel{(i)}{{\bchi}}$ are zero. The
background stress tensor ${\bsi}_{{\rm bg}}$ is also given by
(3.8), but the corresponding stress potential is labelled
appropriately: $\stackrel{(i)}{{\bchi}}_{{\rm bg}}$.

\subsubsection{The first order}

In the first order, only the stress potential
$\stackrel{(1)}{\phi}$ (3.11) is non-zero for the screw
dislocation. The corresponding governing equation appears from
(3.9) as follows \cite{mal}:
$$
(\Dl\,-\,\kappa^2)\,\bigl(\stackrel{(1)}{\phi}
  \,-\,\stackrel{(1)}{\phi}_{{\rm bg}}\bigr)
  \,=\,b\,\dl^{(2)}(x)\,,
\eqno(3.12)
$$
where $\stackrel{(1)}{\phi}_{{\rm bg}} \equiv (-b/2\pi) \log\rho$
is the background stress potential, and $b$ is the Burgers vector
length. The Burgers vector is parallel to the line of the screw
dislocation, and its length is the perturbative expansion
parameter. The axial symmetry of the final state stimulates the
usage of the cylindric coordinates $\rho$, $\ph$, and $z$ instead
of $\{ x^a \}$ ($\rho$ and $\ph$ are chosen in the $(x^1,
x^2)$-plane and $z\equiv x_3$). We obtain from (3.12) the stress
potential of the first order:
$$
\stackrel{(1)}{\phi}\,=\,\stackrel{(1)}{\phi}_{{\rm bg}}
\,-\,f_S\,,\qquad f_S\,\equiv\,(b/2\pi) K_0 (\kappa\rho)\,,
\eqno(3.13)
$$
where $K_0$ is the modified Bessel function \cite{wat}.

The stress $\si_{\phi z}$ is non-trivial for the modified screw
dislocation in the first order, and we express it, using (3.8) and
(3.11):
$$
\si_{\phi z}\,=\,-\,\mu\,\cd_\rho\stackrel{(1)}{\phi}\,=\, \frac{b
\mu}{2\pi}\,\rho^{\1} \bigl(1-\kappa\rho K_1(\kappa\rho)\bigr)\,.
\eqno(3.14)
$$
Equation (3.14) demonstrates a core region at
$\rho\,{\stackrel{<}{_\sim}}\,\kappa^{\1}\,$: the gauge correction
to the classical long-ranged law $1/\rho$ is exponentially small
outside this region. Inside it, the law $1/\rho$ is replaced by
another non-singular one. More detailed information on the
numerical behaviour of (3.14) (including a treatment of
$\kappa^{-1}$ in terms of interatomic spacing) can be found in
\cite{ed3} and \cite{laz2}. In the present paper, it is assumed
that $b= O (\kappa^{-1})$. Solution (3.14) is in agreement with
\cite{ed3}, \cite{laz2} (the translational gauging), \cite{cem}
(the non-local elasticity) and \cite{g1}, \cite{g4}, and
\cite{laz5} (the strain gradient elasticity). This is because the
Helmholtz-type governing equations similar to (3.12) are essential
in the approaches mentioned.

Certain analogies in the structure of the tensor laws governing
the electrostatic of dielectrics and the elastostatic problems
have been discussed in \cite{hoen}. It looks hopeful that the
short-ranged constituent of the first order solution presented
here is also falling into the class of problems considered in
\cite{hoen}. In addition, within an independent framework though,
strongly localized stress potentials turn out to be also
responsible for the Debye-like screening effects in the
dislocation arrangements \cite{groma}.

\subsubsection{The second order}

In the second order, from (3.9) we obtain the following gauge
equations \footnote{Henceforth, ${\bf Q}$ and ${\bPsi}$ are used
without the superscript.}:
$$
\begin{array}{rcl}
\displaystyle{
-\cd^2_{2 2}\Bigl[(1-a)p\,+\,a \Dl f\,-\,
      \kappa^2(f-f_{{\rm bg}})\Bigr]}\,&=&\,2\mu
          \Bigl[Q_{1 1}\,+\,\cd^2_{2 2}\Psi_{3 3}\Bigr]\,,
          \\[0.5cm]
\displaystyle{
-\cd^2_{1 1}\Bigl[(1-a)p\,+\,a \Dl f\,-\,
      \kappa^2(f-f_{{\rm bg}})\Bigr]}\,&=&\,2\mu
          \Bigl[Q_{2 2}\,+\,\cd^2_{1 1}\Psi_{3 3}\Bigr]\,,
          \\[0.5cm]
\displaystyle{
\cd^2_{1 2}\Bigl[(1-a)p\,+\,a \Dl f\,-\,
      \kappa^2(f-f_{{\rm bg}})\Bigr]}\,&=&\,-2\mu\,
          \cd^2_{1 2}\Psi_{3 3}\,,
\end{array}
\eqno(3.15.1)
$$
$$
\begin{array}{rcl}
\displaystyle{
(1-a)\Dl\Dl f\,+\,a \Dl p}&-&
      \kappa^2(p-p_{{\rm bg}}) \\[0.5cm] &=&2\mu
          \Bigl[Q_{3 3}\,-\,\cd^2_{1 2}
\bigl(\Psi_{1 2}\,+\,\Psi_{2 1}\bigr)\,+\,\cd^2_{1 1}\Psi_{2 2}
\,+\,\cd^2_{2 2}\Psi_{1 1}\Bigr]\,.
\end{array}
\eqno(3.15.2)
$$
Equations (3.15) describe the stress potentials of second order
$f$ and $p$ (3.11). In addition, there are equations to determine
$\stackrel{(2)}{\phi}$. But since $\stackrel{(2)}{\phi}_{{\rm
bg}}$ is zero \cite{pfl}, we put consistently
$\stackrel{(2)}{\phi}\equiv 0$. Further, it is necessary to
express ${\bf Q}$ in (3.3.3) and ${\bPsi}$ in (3.6) in terms of
the first order solution (3.13). Taking $\cd_3 = 0$ into account
and using ${e_{a b}}^c$ (2.9.2), we obtain the following non-zero
components:
$$
Q_{1 1}\,=\,Q_{2
2}\,=\,\frac{\bigl(\Dl\stackrel{(1)}{\phi}\bigr)^2}4\,,
\eqno(3.16.1)
$$
$$
Q_{3 3}\,=\,\displaystyle{ \Bigl(\cd^2_{1
2}\stackrel{(1)}{\phi}\Bigr)^2 \,-\,\cd^2_{1
1}\stackrel{(1)}{\phi}\, \cd^2_{2
2}\stackrel{(1)}{\phi}\,+\,\frac{\bigl(\Dl\stackrel{(1)}{\phi}\bigr)^2}4\,.}
\eqno(3.16.2)
$$
In turn, the corresponding components of ${\bPsi}$ are expressed
standardly (see \cite{pfl} and \cite{gair}). Using (3.16.2), we
obtain the combination that we are interested in:
$$
\begin{array}{l}
Q_{3 3}\,-\,2 \cd^2_{1 2} \Psi_{1 2}\,+\,\cd^2_{1 1}\Psi_{2 2}
\,+\,\cd^2_{2 2}\Psi_{1 1} \\[0.5cm]
\qquad \qquad \,=\, \displaystyle{ \Dl\Psi_{3 3}\,+\,
\bigl(1-2\mu^2 C_7\bigr) \,\Phi\, +\, \frac{
\bigl(\Dl\stackrel{(1)}{\phi}\bigr)^2}{4}}\,,
\end{array}
\eqno(3.17)
$$
where
$$
\begin{array}{l}
 \displaystyle{\Psi_{3 3}\,\equiv\,
-\,\mu^2\,C_3\Bigl(\bigl(\cd_1\stackrel{(1)}{\phi}\bigr)^2\,+\,
\bigl(\cd_2\stackrel{(1)}{\phi}\bigr)^2\Bigr) }\,,\\[0.5cm]
\Phi\,\equiv\,\Bigl(\cd^2_{1 2}\stackrel{(1)}{\phi}\Bigr)^2
\,-\,\cd^2_{1 1}\stackrel{(1)}{\phi}\, \cd^2_{2
2}\stackrel{(1)}{\phi}\,.
\end{array}
\eqno(3.18)
$$

Equations (3.16) and (3.17) look unconventionally owing to the
presence of $(\Dl\stackrel{(1)}{\phi})^2/4$. It is just the
dependence of ${\bf Q}$ on the torsion that results in the absence
of the contribution $(\Dl\stackrel{(1)}{\phi})^2/4$ in the
components $Q_{1 1}$, $Q_{2 2}$, and $Q_{3 3}$ obtained in
\cite{pfl}. Recall that the conventional incompatibility law
requires the vanishing of the Einstein tensor calculated by means
of the Riemann--Cartan curvature \cite{lh}. This latter includes
(see \cite{klein}), in addition to the Riemann--Christoffel part
(2.8), a contribution owing to the torsion tensor. The latter is
identified as the dislocation density. That is, the dislocation
density contributes into the driving source of the incompatibility
law. On the contrary, according to \cite{ed3}, \cite{mal}, and
\cite{laz2}, a stress field owing to the background defect is
chosen first. Since the torsion is not considered as an
independent variable, $Q_{1 1}$, $Q_{2 2}$, and $Q_{3 3}$ acquire
the form (3.16).

Equations (3.15) allow for a correct limit to their classical
version, which is respected by the background stress potentials of
second order $f_{{\rm bg}}$ and $p_{{\rm bg}}$. Let us pass to the
classically-looking $Q_{1 1}$, $Q_{2 2}$, and $Q_{1 2}$:
$$
Q_{1 1}\,=\,\cd^2_{2 2} Q^{\prime}\,,\quad Q_{2 2}\,=\,\cd^2_{1 1}
Q^{\prime}\,,\quad Q_{1 2}\,=\,-\cd^2_{1 2} Q^{\prime}\,.
\eqno(3.19)
$$
A specific value of the constant $Q^\prime$ arises from the
requirement that $\stackrel{(2)}{\si}_{3 3}$ averaged over the
cylinder's cross-section is zero (see \cite{pfl} and \cite{gair}).
Then, the equations that govern $f_{{\rm bg}}$ and $p_{{\rm bg}}$
take in our notation the form ($\kappa\to 0$)
$$
\begin{array}{l}
\Dl\Dl f_{{\rm bg}}\,=\, k\,\Bigl[\Dl (\Psi^{{\rm bg}}_{3
3}+Q^{\prime})\,+\,(1-a)\,(1\,-\,2\mu^2 C_7)\,\Phi_{{\rm
bg}}\Bigr]\,,
\\[0.5cm]
(1-a)\,p_{{\rm bg}}\,+\,a\,\Dl f_{{\rm bg}}\,=\,
-2\mu\,\bigl(\Psi^{{\rm bg}}_{3 3}+Q^\prime\bigr)\,,
\end{array}
\eqno(3.20)
$$
where
$$
k\,\equiv\,\frac{2\mu}{1-2a}\,=\,2\mu\,\frac{1+\nu}{1-\nu}\,.
\eqno(3.21)
$$
Moreover, $\Psi^{{\rm bg}}_{3 3}$ and $\Phi_{{\rm bg}}$ are given
by (3.18), provided that $\stackrel{(1)}{\phi}$ is replaced by
$\stackrel{(1)}{\phi}_{{\rm bg}}$.

Let us turn to (3.16). In the present approach,
$\bigl(\Dl\stackrel{(1)}{\phi}\bigr)^2$ is the square of the
density distribution of the modified screw dislocation (see
\cite{ed3}, \cite{mal}, and \cite{laz2}):
$$
\bigl(\Dl\stackrel{(1)}{\phi}\bigr)^2\,=\,
\Bigl(\frac{b}{2\pi}\,\kappa^2\,K_0(\kappa\rho)\Bigr)^2\,.
\eqno(3.22)
$$
For large distances, $K_0(\kappa\rho)$ decays exponentially, and
$\bigl(\Dl\stackrel{(1)}{\phi}\bigr)^2$ given by (3.22) can
approximately be replaced by zero. Therefore, the components
(3.16) take, also approximately, their conventional form. Then,
three equations (3.15.1) reduce to a single one. However,
$\Dl\stackrel{(1)}{\phi}$ is not negligible for sufficiently
moderate $\kappa\rho$, and some care inside the core region
$\rho\,{\stackrel{<}{_\sim}}\,\kappa^{-1}$ is required. In other
words, this might be an indication for extension of the geometric
framework by means of the differential-geometric torsion. But this
would, in turn, imply that the teleparallel description is
abandoned.

Instead, we shall establish an ``effective'' picture, which should
be viewed as still contained within the teleparallel framework.
However, this description is expected to incorporate certain
features of the approach extended differential-geometrically.
Indeed, an independent non-trivial contribution owing to the
torsion would lead to the following equation, instead of (3.12):
$$
\Dl\stackrel{(1)}{\phi}_\cT\,=\,
\kappa^2\bigl(\stackrel{(1)}{\phi}_\cT
\,-\,\stackrel{(1)}{\phi}_{{\rm bg}}\bigr)\,+\,\cT\,. \eqno(3.23)
$$
Here $\cT$ is a suitable density caused by the non-triviality of
the torsion components, say, $T^3_{1 2}=-T^3_{2 1}$ (this is just
appropriate for the screw dislocation along $Oz$, \cite{lh}). The
notation $\stackrel{(1)}{\phi}_\cT$ implies the corresponding
solution at a given $\cT$. Being considered as an additional
source of incompatibility, $\cT$ can be assigned to possess near
$Oz$ a series expansion with arbitrary, though adjustable,
coefficients. As a result, $\stackrel{(1)}{\phi}_\cT$ can also
acquire, in comparison with $\stackrel{(1)}{\phi}$ (3.13), certain
modifications near $\rho=0$.

Therefore, $\bigl(\Dl\stackrel{(1)}{\phi}\bigr)^2$ is replaced, on
an extended treatment, by
$\bigr(\Dl\stackrel{(1)}{\phi}_\cT\bigl)^2$. This can be
effectively accounted for in the present framework as well. Let us
introduce a piece-wisely constant density to ``regularize''
$\bigl(\Dl\stackrel{(1)}{\phi}\bigr)^2$ as follows \cite{ml}:
$$
\bigl(\Dl\stackrel{(1)}{\phi}\bigr)^2 \,\sim\, \Bigl( \frac{b}{\pi
\rho^2_*}\Bigr)^2\,h_{[0,\,\rho_*]}(\rho)\,, \eqno(3.24)
$$
where $h_{[0,\,\rho_*]}(\rho)$ is unity for $\rho\in [ 0, \rho_*]$
(i.e., within a disc) or zero otherwise. It is crucial that (3.24)
should not be treated as a replacement to be iteratively improved.
Let us use (3.24) in (3.16.1). Then, an analogue of (3.19) is
valid, where the corresponding ``potential'' ${\w g}$ is present:
$$
{\w g}\,=\,\frac{b^2}{8\pi^2\rho^2_*}\,
\Bigl(\frac{\rho^2}{\rho^2_*}\,-\,1\Bigr)\,
h_{[0,\,\rho_*]}(\rho)\,+\,{\cC}\,. \eqno(3.25)
$$
The constants ${\cC}$ and $\rho_*$ should be specified later. Both
$\displaystyle{\frac{b}{2\pi} \kappa^2 K_0(\kappa\rho)}$ and
$\displaystyle{\frac{b}{\pi \rho^2_*} h_{[0,\,\rho_*]}(\rho)}$ are
$\dl$-like for $\kappa$ and $1/\rho_*$ large enough, respectively.
Considered as the surface densities, they are properly normalized.
Therefore, the replacement (3.24) looks better, provided that its
$\dl$-like character is sharper.

Using (3.19), where ${\w g}$ given by (3.25) is substituted, from
(3.15) we obtain the following equations \cite{ml}:
$$
\displaystyle{(\Dl\,-\,\kappa^2)\,\Bigl(\Dl\,+\,
\frac{\kappa^2}{1-2a}\Bigr)\,(f-f_{{\rm bg}})\,=\,k\,{\cR}}\,,
\eqno(3.26.1)
$$
$$
\displaystyle{ p\,=\,-\frac{a}{1-a}\,\Dl f\,-\,
\frac{2\mu}{1-a}\,(\Psi_{3 3}\,+\,{\w g})}
\,+\,\frac{\kappa^2}{1-a} \,(f-f_{{\rm bg}})\,. \eqno(3.26.2)
$$
Here, $f_{{\rm bg}}$ respects the first equation (3.20), and $k$
is given by (3.21). The driving source in (3.26.1) is defined as
follows:
$$
\begin{array}{rcl}
{\cR} &\equiv& (\Dl\,-\,\kappa^2) \bigl(\Psi_{3 3}\,+\,{\w
g}\,-\,\Psi^{{\rm bg}}_{3 3}
\,-\,Q^\prime\bigr) \\[0.5cm]
&+&\, \displaystyle{ (1-a)\bigl(1\,-\,2\mu^2 C_7\bigr)
\bigl(\Phi\,-\,\Phi_{{\rm bg}}\bigr)\,-\,\frac{1-a}4
\bigl(\Dl\stackrel{(1)}{\phi}\bigr)^2}\,.
\end{array}
\eqno(3.27)
$$
It is assumed that $\Psi_{3 3}$ and $\Phi$ (or, analogously,
$\Psi^{{\rm bg}}_{3 3}$ and $\Phi_{{\rm bg}}$) in (3.27) are given
by (3.18) where $\stackrel{(1)}{\phi}$ (or, correspondingly,
$\stackrel{(1)}{\phi}_{{\rm bg}}$) is substituted appropriately.
In addition, $\bigl(\Dl\stackrel{(1)}{\phi}\bigr)^2$ is kept in
(3.27) only formally, as a symbol. The first equation in (3.26.1)
(would be called, after \cite{lam1} and \cite{lam2}, the
non-homogeneous {\it bi-Helmholtz equation} \footnote{ See
\cite{lam1} and \cite{lam2} for appropriate refs. and more
examples of non-homogeneous bi-Helmholtz equations. A $\dl$-driven
differential equation analogous to (3.26.1) has already appeared
in \cite{mal} as well.}) is just to be solved in what follows to
determine the stress function $f$. Equation (3.26.2) is needed
only to express the stress component $\si_{z z} = p$.

The governing equations (3.26) are essentially similar to
equations that should be expected under a consideration extended
by means of the torsion (see also explanations in Section 5.3).
The actual derivation of (3.26) incorporates (3.16) and (3.17),
which are subject to (3.24). Therefore, the teleparallel framework
is valid only approximately. However, the free parameter $\rho_*$
is coming to play by means of (3.24), and this supports the
similarity noted. Indeed, the arbitrariness owing to $\rho_*$ is
reminiscent of a freedom caused by the non-triviality of
appropriate torsion. Technical complexities, which might appear in
the description based on (3.23), should be avoided now, since they
deserve separate studying. Therefore, the strategy based on (3.24)
looks appropriate. Equations (3.26) lead eventually to a
self-consistent stress distribution without an artificial cut-off
at the core.

\subsection{The driving source of the gauge equation}

Before solving (3.26.1), simplifications for ${\cR}$ given by
(3.27) are in order. It is convenient to investigate ${\cR}$
multiplied by $\rho^2$. We use (3.13) and (3.18) and keep the same
notation ${\cR}$ after the multiplication. After re-scaling the
radial coordinate $\kappa\rho\mapsto s$, we obtain approximately
(see \cite{ml}):
$$
{\cR}\,\approx\,s^2 w(s)\,+\,\sum\limits_{a=1}^3 {\cR}_{a}\,,
\eqno(3.28)
$$
where
$$
\begin{array}{rcl}
X^{-2}\,{\cR}_{1}&=&\displaystyle{ \bigl({\tilde c}\,+\,c
s^2\bigr)\,K_1(s)\, \Bigl(K_1(s)\,-\,\frac 2s\Bigr)}\,,
\\[0.5cm]
X^{-2}\,{\cR}_2&\approx& -{\tilde c}s K_1(s)\Dl_s {\w\phi}
\,\sim\,\displaystyle{ \frac{2{ \tilde c}}{s^2_*}\,s K_1(s)\,{\w
h}_{\,[0,\, s_*]}(s)} \,,
\\[0.5cm]
X^{-2}\,{\cR}_3&\approx & \displaystyle{ -2 c\,s^2
K_1(s)\frac{d}{d s}\Bigl(\Dl_s {\w\phi}\Bigr)} \,=\,-2 c\,\Bigl(s
K_1(s)\Bigr)^2 \,.
\end{array}
\eqno(3.29)
$$
The following notation is accepted in (3.28) and (3.29):
$$
\begin{array}{l}
\displaystyle{ {\tilde c}\,\equiv\,(1-a)(1-2\mu^2 C_7)\,-\,4c\,,
\qquad \qquad c\equiv \mu^2 C_3
}\,,\\[0.5cm]
s^2\,w(s) \,\equiv\, \displaystyle{
X^2\,\frac{s^2}{s^4_*}\Bigl[(1+a)\,-\,
      \frac12\bigl(s^2-s^2_*\bigr)\Bigr]\,
           {\w h}_{\,[0,\, s_*]}(s)}\,,
\end{array}
\eqno(3.30)
$$
where $a$ and $1-a$ are given by (3.10), $s_*\equiv\kappa\rho_*$,
and ${\w h}_{\,[0,\, s_*]}(s)$ is unity at $s\in [0,\, s_*]$ or
zero otherwise (we use ${\cC} = Q^{\prime}$ in (3.25)). Further,
${\w\phi}$ in (3.29) implies $\displaystyle{\stackrel{(1)}{\phi}}$
with removed multiple $\displaystyle{\frac{b}{2\pi}}$. The factor
$X^2\equiv\displaystyle{\Bigl(\frac{b\,\kappa}{2 \pi}\Bigr)^2}$
just points out the fact that the driving source is quadratic in
$b$. Moreover, $\Dl_s$ is $s^{\1}\displaystyle{ \frac{d}{d s}
\Bigl(s\,\frac{d}{d s}\Bigr)}$, and $K_0$ and $K_1$ are the
modified Bessel functions \cite{wat}.

The following comments on (3.28) and (3.29) are in order.
Classically, $\Dl \stackrel{(1)}{\phi}_{{\rm bg}}$ is the Dirac's
delta-function, and its square would occur instead of (3.22). The
equation for the corresponding stress function $f_{\rm bg}$ is
conventionally considered for strictly positive $\rho$. Then, the
corresponding driving source is not influenced by $\Dl
\stackrel{(1)}{\phi}_{{\rm bg}}$. Therefore, it seems
inappropriate to account for, at the same footing with (3.24), the
contributions expressible in ${\cR}$ by the differences $\Dl
\stackrel{(1)}{\phi} - \Dl \stackrel{(1)}{\phi}_{{\rm bg}}$ or
$\cd_\rho(\Dl \stackrel{(1)}{\phi} - \Dl
\stackrel{(1)}{\phi}_{{\rm bg}})$. To put it differently, the
regularization allowed classically ``persists'' into the
non-conventional approach, and (3.28) and (3.29) take their actual
form.

The asymptotic properties of ${\cR}$ (3.28) are as follows. First,
$\cR$ is well localized since it contains the modified Bessel
functions. At small $s$, we expand
$$
\begin{array}{rcl}
{\cR}(s)&\simeq & p_1\,s^{-2}\,+\,p_2\,\log s
\,+\,p_3\\[0.5cm]
&+&s^2\,\bigl(p_4\,\log^2 s\,+\,p_5\,\log
s\,+\,p_6\bigr)\,+\,o(s^2)\,,
\end{array}
\eqno(3.31)
$$
where the numerical coefficients $p_1, p_2, \dots, p_6$ are
$$
\begin{array}{rcl}
p_1&=&-X^2\,{\tilde c}\,, \qquad \qquad \qquad p_2\,=\,0\,,
\\[0.5cm]
p_3&=&\displaystyle{ X^2\,\Bigl(\frac{2{\tilde c}}{s^2_*} - \al
c\Bigr)}\,, \qquad \displaystyle{
p_4\,=\,X^2\,\frac{{\tilde c}}4 }\,, \\[0.5cm]
p_5&=& \displaystyle{ X^2\Bigl(\frac{{\tilde c}}{s^2_*}\,-\,
\frac{{\tilde c}}4\,\Bigl(1-2\log \frac \ga{2}\Bigr)\,+\,(1-\al)\,
c\Bigr)}\,,
\\[0.5cm]
p_6&=&w(0)\,+\, \displaystyle{ X^2\,\Bigl(1-2\log \frac
\ga{2}\Bigr)\,\Bigl( \Bigl(1-2\log \frac
\ga{2}\,-\,\frac{8}{s^2_*}\Bigr)\, \frac{{\tilde c}}{16}\,+\,(\al
- 1)\,\frac{c}2 \Bigr)}\,,
\end{array}
\eqno(3.32)
$$
with $\al=1$ or $\al=3$ in $p_3, p_5$, and $p_6$. Indeed, the
calculation of ${\cR}_3$ requires, as soon as (3.24) is imposed,
the differentiation of the step-function. This would result in
singularities in the driving source. Our regularized scheme can
thus be destroyed. We use $p_3, p_5$, and $p_6$ with $\al=1$, when
${\cR}_3$ is simply omitted to express the neglect of the
corresponding differentiation. Otherwise (for comparison), $\al=3$
when ${\cR}_3$ is calculated according to (3.29). However, $\al$
is left unspecified to give an indication of both possibilities.
Moreover, in spite of $p_2\equiv 0$, it is instructive to keep the
corresponding term in (3.31). It is to be equated to zero at the
end.

\section{Solution of the gauge equation}

We are going to solve (3.26.1) in two steps \cite{ml}. First, we
consider the non-homogeneous Bessel equation
$$
\Bigl[s^2\frac{d^2}{d s^2}\,+ \,s \frac{d}{d s}\,-\,s^2\Bigr]
G(s)\,=\,{\cR}(s)\,. \eqno(4.1)
$$
The method of variation of parameters \cite{mag} provides a
general solution to (4.1). We choose the modified Bessel functions
$I_0$ and $K_0$ \cite{wat} as two linearly independent solutions
of the associated homogeneous Bessel equation. Solution to (4.1)
appears as follows:
$$
\begin{array}{rcl}
G(s)&=& \displaystyle{\lim_{\epsilon\to 0}}\,\Bigg[\displaystyle{
I_0(s)\,\Bigl(A(\epsilon)\,+\,
\int\limits^s_\epsilon K_0(t) {\cR}(t)\,\frac{dt}{t}\Bigr)}\\
[0.5cm] &+&\displaystyle{ K_0(s)\,\Bigl(B(\epsilon)\,-\,
\int\limits^s_\epsilon I_0(t)
{\cR}(t)\,\frac{dt}{t}\Bigr)\Bigg]}\,,
\end{array}
\eqno(4.2.1)
$$
where
$$
\begin{array}{l}
A(\epsilon)\,\equiv\,-\,\displaystyle{ \int\limits^\infty_\epsilon
K_0(t) {\cR}(t)\,\frac{dt}{t}}\,,\\ [0.5cm]
B(\epsilon)\,\equiv\,{\rm
const}\,+\,\displaystyle{\int\limits^1_\epsilon
\biggl(\frac{p_1}{t^3}\,+\,\frac{p_2}{t}\,\log t\,+\,
\Bigl(\frac{p_1}4\,+\,p_3\Bigr)\frac 1t\biggr)\,dt}\,.
\end{array}
\eqno(4.2.2)
$$

The behaviour of $t^{-1} I_0(t) {\cR}(t)$ and $t^{-1} K_0(t)
{\cR}(t)$ for small or large $t$ is important for justification of
(4.2). When $t$ is small, the regularization in $G(s)$ is ensured
by means of the specially constructed $A(\epsilon)$ and
$B(\epsilon)$ from (4.2.2) (see Appendix A). When $t$ is large,
$K_0(t) {\cR}(t)$ decays as $e^{-2t}$, whereas $I_0(t) {\cR}(t)$
behaves, mainly, as a constant. Therefore, the infinity as an
upper bound in $A(\epsilon)$ given by (4.2.2) is appropriate. The
choice of the integration bounds implies that $G(s)$ decays
exponentially for large $s$. The important arbitrariness in
$B(\epsilon)$ from (4.2.2) should be stressed, which is due to the
additive constant denoted as $const$.

Using Appendix $A$, it is straightforward to expand $G(s)$ for
small $s$:
$$
G(s)\simeq  q_1\,s^{-2}\, +\,\displaystyle{\sum_{i=0}^3
q_{5-i}\log^i s\,+\,s^2\,\sum_{i=0}^3 q_{9-i}\log^i s\,+\,\dots
}\,. \eqno(4.3)
$$
The numerical coefficients in (4.3) are expressed in terms of the
parameters (3.32) as follows:
$$
\begin{array}{l}
\displaystyle{ q_1\,=\,\frac{p_1}4\,,\quad \qquad
q_2\,=\,\frac{p_2}6\,,\qquad \quad
q_3\,=\,\frac{p_1}8\,+\,\frac{p_3}2\,,}
\\[0.5cm]
\displaystyle{ q_4\,=\,-\,{\rm const}\,+\,\frac{3 p_1}4\,, \qquad
q_5\,=\,-\,{\rm const}\times\log\frac{\ga}2\,-\, {\cal
I}_K\,-\,\frac{p_1}{16}\,,}
\\[0.5cm]
\displaystyle{ q_6\,=\,\frac{p_2}{24}\,,\qquad \quad
q_7\,=\,\frac{p_1}{32} \,+\,\frac{p_3-p_2}8\,+\,\frac{p_4}4\,,}
\end{array}
\eqno(4.4)
$$
where ${\cal I}_K$ is given by (A8) in Appendix A, and ${\it
const}$ in $q_4$ and $q_5$ is introduced by (4.2.2). The
coefficients $q_8$ and $q_9$ in (4.3) are practically too
complicated and are not of importance below.

As a second step, we express the difference $f - f_{\rm bg}\equiv
k\,{\cF}$, where ${\cF}$ respects the non-homogeneous Bessel
equation
$$
\Bigl[s^2\frac{d^2}{d s^2}\,+ \,s \frac{d}{d s}\,+\,s^2\Bigr]
{\cF}\Bigl(\frac{s}{\cN}\Bigr)\,=\,
\frac{s^2}{{\cN}^2}\,G\,\Bigl(\frac{\kappa}{{\cN}}\,s\Bigr)\,.
\eqno(4.5)
$$
The modified stress potential of the second order appears as $f =
f_{\rm bg} + k\,{\cF}$. For convenience, (4.5) is also written in
terms of the new variable $s$, which is however defined
differently: $s\equiv {\cN}\rho$, where
${\cN}^2\equiv\displaystyle{\frac{\kappa^2}{1-2a}}$. We choose
$Y_0$ and $J_0$ as the fundamental solutions of the homogeneous
Bessel equation associated with (4.5) and obtain
$$
\begin{array}{l}
{\cF}(\rho)\,=\,C\,{\w Y}_0({\cN}\rho)\,+\,D\,J_0({\cN}\rho)
\,+\,I_{{\cF}}(\rho)\,,\\[0.5cm]
I_{{\cF}}(\rho)\,\equiv\,
\displaystyle{
J_0({\cN}\rho)\,\int\limits^\infty_\rho {\w Y}_0({\cN} t)
G(\kappa t)\,t\,dt}\,
-\,\displaystyle{{\w Y}_0({\cN}\rho)\,
\int\limits^\infty_\rho J_0({\cN}t) G(\kappa t)\,t\,dt}\,,
\end{array}
\eqno(4.6)
$$
where $G(\cdot)$ is given by (4.2), and ${\w Y}_0(s)\equiv
(\pi/2)Y_0(s)$. Here the integrals are convergent at their upper
bounds.

In the preceding consideration (see \cite{ml}), the choice of
$C\ne 0$ and $D=0$ has been made in ${\cF}(\rho)$ from (4.6).
However, in this case the stress field looks non-conventional for
sufficiently large distances. In the present paper, we put $C$ and
$D$ equal to zero, and thus ${\cF}(\rho)=I_{{\cF}}(\rho)$. The
integral $I_{{\cF}}(\rho)$ decays exponentially for large $\rho$,
while for small ${\cN}\rho$ we obtain (Appendix $B$)
$$
{\cF}(\rho)\,\simeq\,\displaystyle{
r_0\,+\,r_1\,\log\rho\,+\,r_2\,\log^2\rho\,+\,
\rho^2\,\sum_{i=0}^3 r_{6-i}\,\log^i\rho }\,+\,r_7\,\rho^4\log^3
\rho\,+\,\dots\,, \eqno(4.7)
$$
where
$$
\begin{array}{l}
\displaystyle{ r_0\,=\,-\,{\w {\cal
I}_Y}\,+\,\log\Bigl(\frac{\ga}2{\cN}\Bigr) \,{\w {\cal
I}_J}\,,\qquad \quad r_1\,=\,{\w {\cal I}_J}\,,\qquad \quad
r_2\,=\,
\frac{p_1}{8\kappa^2}}\,,\\[0.5cm]
\displaystyle{ r_3\,=\,\frac{p_2}{24}\,,\qquad r_4\,=\,
\Bigl(1\,-\,\frac{{\cN}^2}{\kappa^2} \Bigr)\,
\frac{p_1}{32}\,-\,(1\,-\,\log\kappa)\frac{p_2}8\,
+\,\frac{p_3}8 \,,}\\[0.5cm]
\displaystyle{ r_7\,=\,(\kappa^2-{\cN}^2) \frac{p_2}{384}}\,,
\end{array}
\eqno(4.8)
$$
and $p_1, p_2$, and $p_3$ are given by (3.32). The coefficients
$r_5$ and $r_6$ are given by (B6)--(B11), and ${\w {\cal I}_Y}$
and ${\w {\cal I}_J}$ are defined by (B10) in Appendix B.

\section{The stress tensor}
\subsection{The components ${\si}_{\rho \rho}$ and
${\si}_{\phi \phi}$ }

Therefore the modified stress potential of second order $f$ takes
the form
$$
f\,=\,f_{{\rm bg}}\,+\,k I_{{\cF}}\,,
\eqno(5.1.1)
$$
$$
f_{{\rm bg}}\,=\,-\,\nt\,\log^2\rho
\,+\,d_1\,\rho^2\,+\,d_2\,\log \rho\,, \qquad
\nt\,\equiv\,k\,\frac{p_1}{8\kappa^2}\,,
\eqno(5.1.2)
$$
where $I_{\cF}$ is given by (4.6), and the stress potential
$f_{{\rm bg}}$ is in full agreement with \cite{pfl}. The potential
$f_{{\rm bg}}$ respects the following equation, which results from
(3.20) for $\rho\ne 0$:
$$
\Dl \Dl f_{{\rm bg}}\,=\,\frac{-8 \nt}{\rho^4}\,. \eqno(5.2)
$$

Taking into account (3.8) and (3.11), we represent the stress
tensor of the second order in the cylindrical coordinates as
follows:
$$
\stackrel{(2)}{\si}_{\rho \rho}\,=\,
-\,\frac{1}{\rho}\,\frac{d}{d \rho}\,f\,,\qquad
\stackrel{(2)}{\si}_{\phi \phi}\,=\,
-\,\frac{d^2}{d \rho^2}\,f\,,\qquad
\stackrel{(2)}{\si}_{z z}\,=\,p\,,
\eqno(5.3)
$$
where $f$ from (5.1) and $p$ from (3.26.2) are substituted. The
other components of $\stackrel{(2)}{{\bsi}}$ are zero. In the
classical problem, we use (5.3) with $f_{{\rm bg}}$ from (5.1.2)
and obtain
$$
\begin{array}{rcl}
(\stackrel{(2)}{\si}_{{\rm bg}})_{\rho \rho}
     & = &\displaystyle{ 2\nt\,\frac{\log \rho}{\rho^2}}\,-\,
     \frac{d_2}{\rho^2}\,-\,2 d_1 \,,
\\ [0.5cm]
(\stackrel{(2)}{\si}_{{\rm bg}})_{\phi \phi} & = & \displaystyle{-
2 \nt\,\frac{\log \rho}{\rho^2}}\,+\,\frac{2 \nt +
d_2}{\rho^2}\,-\,2 d_1\,.
\end{array}
\eqno(5.4)
$$
The free parameters $d_1$ and $d_2$ in (5.4) are given by the
vanishing of $(\stackrel{(2)}{\si}_{{\rm bg}})_{\rho \rho}$ on the
boundaries of the hollow cylinder, e.g., $\rho=\rho_\c$ and
$\rho=\rho_\e > \rho_\c$:
$$
(\stackrel{(2)}{\si}_{{\rm bg}})_{\rho \rho} \Big
|_{\rho=\rho_\e}\,=\, (\stackrel{(2)}{\si}_{{\rm bg}})_{\rho
\rho}\Big |_{\rho=\rho_\c}\,=\,0\,. \eqno(5.5)
$$

Substitution of (5.4) into (5.5) allows us to define $d_1$ and
$d_2$. For our purposes, it is more appropriate to assume that
$\rho_\e$ is essentially greater than $\rho_\c$ ($\rho_\e/\rho$ is
moderate). We approximately obtain
$$
\displaystyle{ d_1\,\approx\,\frac{\nt}{\rho^2_\e}\,
\log\frac{\rho_\e}{\rho_\c}\,,\qquad\, d_2\,\approx\,2
\nt\,\log\rho_\c }\,, \eqno(5.6)
$$
and
$$
\begin{array}{rcl}
(\stackrel{(2)}{\si}_{{\rm bg}})_{\rho \rho} &\approx&
\displaystyle{ \frac{2 \nt}{\rho^2}
\biggl(\log\frac{\rho}{\rho_\c}\,-\,
\Bigl(\frac{\rho}{\rho_\e}\Bigr)^2\log\frac{\rho_\e}{\rho_\c}
\biggr) }\,,\\ [0.5cm]
(\stackrel{(2)}{\si}_{{\rm bg}})_{\phi
\phi} &\approx& \displaystyle{ \frac{2 \nt}{\rho^2}
\biggl(1\,-\,\log\frac{\rho}{\rho_\c}\,-\,
\Bigl(\frac{\rho}{\rho_\e}\Bigr)^2\log\frac{\rho_\e}{\rho_\c}
\biggr) }\,.
\end{array}
\eqno(5.7)
$$
Note that $\nt$ is proportional to $b^2$, and $f_{{\rm bg}}$ from
(5.1.2) is indeed of second order in $b$. From (3.29), (4.2), and
(4.6) it is seen that $I_{\cF}$ is also quadratic in $b$.

Let us turn to the stress components (5.3) with $f$ substituted
from (5.1). The solution in question depends on several free
parameters: $\rho_*$, $const$, $d_1$, and $d_2$ (see (3.25),
(4.2.2), and (5.1.2), respectively). Moreover,
$\stackrel{(2)}{\si}_{z z}$ from (3.26.2) depends on ${\cC}$ ($=
Q^{\prime}$). In order to avoid the arbitrariness, the
asymptotical stresses should be subject to appropriate conditions
at $\kappa\rho\gg 1$ and $\kappa\rho\ll 1$. For instance,
$I_{{\cF}}$ is exponentially small for $\kappa\rho\gg 1$, and
therefore $f\simeq f_{{\rm bg}} (\rho)$. As a result,
$\stackrel{(2)}{\si}_{\rho \rho} \simeq (\stackrel{(2)}{\si}_{{\rm
bg}})_{\rho \rho}$ and $\stackrel{(2)}{\si}_{\phi \phi} \simeq
(\stackrel{(2)}{\si}_{{\rm bg}})_{\phi \phi}$, where the
background stresses are given by (5.4). However, $d_1$ and $d_2$
in the gauge case are defined below.

In the limit $\kappa\rho\ll 1$, we use (4.7) and (5.1) and obtain
$\stackrel{(2)}{\si}_{\rho \rho}$ and $\stackrel{(2)}{\si}_{\phi
\phi}$ as follows:
$$
\begin{array}{rcl}
\stackrel{(2)}{\si}_{\rho \rho}\,-\, (\stackrel{(2)}{\si}_{{\rm
bg}})_{\rho \rho}\,\simeq&-& \displaystyle{ 2
r_2\,\frac{\log\rho}{\rho^2}\,-\,\frac{r_1}{\rho^2}}\,-\,2
r_3\log^3\rho\,-\,
(3 r_3+2 r_4) \log^2\rho \\[0.5cm]
&-&2 (r_4+r_5)\log\rho\,-\,(r_5+2r_6)\,-\, 4 r_7
\rho^2\log^3\rho\,+\,\dots\,,\\ [0.5cm] \stackrel{(2)}{\si}_{\phi
\phi}\,-\,(\stackrel{(2)}{\si}_{{\rm bg}})_{\phi \phi}\,\simeq& &
\displaystyle{ 2 r_2\,\frac{\log\rho}{\rho^2}\,
-\,\frac{2 r_2\,-\,r_1}{\rho^2}} \\
[0.5cm] &-& 2 r_3\log^3\rho\,-\,(9 r_3\,+\,2 r_4) \log^2\rho\,-\,
2 (3 r_3\,+\,3 r_4\,+\,r_5)\log\rho \\[0.5cm]
&-&(3 r_5\,+\,2 r_4\,+\,2 r_6)\,-\,12 r_7
\rho^2\log^3\rho\,+\,\dots \,,
\end{array}
\eqno(5.8)
$$
where the background stresses on the left-hand side are given by
(5.4). Now, $r_1$, $r_2$, ..., and $r_7$ include, for notational
compactness, the multiple $k$ from (5.1.1). The contributions
$\propto\rho^{-2}\log\rho$ disappear on both sides of (5.8) since
$r_2$ is equal to $\nt$. The contributions due to $r_3$ and $r_7$
disappear in (5.8) since $p_2=0$ in (3.31). However, the terms,
which are either divergent or constant as $\rho\to 0$, are still
contained in (5.8).

The number of the free parameters is smaller in comparison with
\cite{ml} since now $C=0$ in (4.6). Therefore, the inappropriate
terms in (5.8) should be handled differently. Recall that the core
region corresponds to $\rho\,{\stackrel{<}{_\sim}}\,\kappa^{-1}$.
Let us introduce a fictitious boundary at $\rho=\rho_\ep$, where
$\rho_\ep\ll {\kappa}^{-1}$. We assume that $\rho$ varies within
the segment $[\rho_\ep, \rho_\e]$, where $\rho_\ep$ is allowed to
be arbitrarily small but strictly non-zero. Then we arrive at the
following equations:
$$
\begin{array}{l}
\displaystyle{ \frac{r_1\,+\,d_2}{\rho_\ep^2}}\,+\,2 r_4
\log^2\rho_\ep \,+\,2
(r_4+r_5)\log\rho_\ep\,+\,r_5\,+\,2(r_6 + d_1)\,=\,0\,,\\
[0.5cm] \displaystyle{ \frac{r_1\,+\,d_2}{\rho_\ep^2}}\,-\,2 r_4
\log^2\rho_\ep \,-\,2 (3\,r_4+r_5)\log\rho_\ep\,-\,3\,r_5\,-\,2
(r_4 + r_6 + d_1)\,=\,0\,,\\
[0.5cm] 2 \nt\log\rho_\e\,-\,d_2\,-\,2 \rho^2_\e\,d_1\,= \,0 \,.
\end{array}
\eqno(5.9)
$$
The first two equations in (5.9) imply that the terms, which are
either divergent or constant in (5.8), vanish at $\rho=\rho_\ep$.
Therefore, $\stackrel{(2)}{\si}_{\rho \rho}$ and
$\stackrel{(2)}{\si}_{\phi \phi}$ at $\rho=\rho_\ep$ are mainly
given by the terms that go to zero as $\rho_\ep\to 0$. The third
equation in (5.9) reads that the radial stress is zero at the
outer boundary $\rho=\rho_\e$. The arbitrariness of $d_1$, $d_2$,
and $const$ can be used to meet (5.9). Since $d_1$, $d_2$, and
$const$ are thus determined (as functions of $\rho_\ep$, in fact),
we can use them in (5.1) and (5.3) to produce the stress
distribution valid on the ring, say, $\cD_{[\rho_\ep, \rho_\e]}$
confined between $\rho=\rho_\ep$ and $\rho=\rho_\e$.

Let us consider an infinite sequence of concentric circles with
monotonically decreasing radii $\rho=\rho_\ep$ ($\rho_\ep$'s
remain however non-zero). Each of the corresponding rings
$\cD_{[\rho_\ep, \rho_\e]}$ is completely covered by a disc with
the same external boundary $\rho=\rho_\e$ but with a puncture at
$\rho=0$. The stress distributions defined on the rings also form
a sequence. It can naturally be considered as converging to an
appropriate solution defined on the punctured disc. Such a limit
is expected to exist, since the boundary values
$\stackrel{(2)}{\si}_{i j}\Big|_{\rho=\rho_\ep}$ go to zero for
decreasing $\rho_\ep$. Therefore, our attention is focussed at the
rings possessing sufficiently small internal radii. Note that even
the definition (4.2) is concerned with a hollow domain possessing
a sufficiently small internal radius. It is just the
regularization in (4.2), which allows us to send this radius to
zero immediately.

To extract the dependence on $const$ in (5.9), let us turn to
$q_4$ and $q_5$ from (4.4). Using Appendix B, one can establish
that $r_1$ in (4.8) and $r_5$, $r_6$ in (B11) depend on $const$ as
follows:
$$
r_1\,\equiv\,-\,{\w C}\,+\,{\w r}_1\,,\quad r_5\,\equiv\,
-\frac{\kappa^2}4\,{\w C}\,+\,{\w r}_5\,,\quad
r_6\,\equiv\,\frac{q\,\kappa^2 }4\,{\w C}\,+\,{\w r}_6\,.
\eqno(5.10)
$$
The following notation is accepted in (5.10):
$$
{\w C}\equiv {\rm const} \times
\displaystyle{\frac{k}{\kappa^2\,+\,{\cN}^2}}\,,\qquad
q\,\equiv\,1\,-\,\log\Bigl(\frac\ga 2\,\kappa\Bigr)\,, \eqno(5.11)
$$
where $\kappa^2=\mu/\ell$ ($\mu$ is the shear modulus, and $\ell$
arises in (2.10)) and $(\kappa/\cN)^2=1-2 a$. Moreover, $a$ and
$k$ are given by (3.10) and (3.21), respectively. We substitute
(5.10) into (5.9) and rewrite the result as a single $3 \times 3$
matrix equation:
$$
\left(
\begin{array}{ccc}
a & 0& 1\\
b & 2 \rho_\ep^2 & 1\\
0 & 2 \rho^2_\e & 1
\end{array}\right)
\left(
\begin{array}{c}
{\w C} \\d_1\\  d_2
\end{array}\right)\,=\,
\left(\begin{array}{c} l_1 \\l_2\\ l_3
\end{array}\right)\,,
\eqno(5.12)
$$
where
$$
\begin{array}{l}
\displaystyle{ a \,\equiv\,- 1\,+\,\frac{(\kappa\rho_\ep)^2}4}
\,,\\[0.5cm]
\displaystyle{ b \,\equiv\,- 1\,-\,\frac{(\kappa\rho_\ep)^2}2
\Bigl(\frac12\,-\,q\,+\,\log\rho_\ep\Bigr)
}\,,\\[0.5cm]
l_1\,\equiv\,2 r_4 \rho_\ep^2 \log\rho_\ep\,+\,(r_4+{\w r}_5)
\rho_\ep^2\,-\,{\w r}_1\,,\\[0.5cm]
l_2\,\equiv\,- 2 r_4 \rho_\ep^2 \log^2\rho_\ep\,-\,2 (r_4+{\w r}_5)
\rho_\ep^2 \log\rho_\ep
\,-\,({\w r}_5 + 2 {\w r}_6)\rho_\ep^2\,-\,{\w r}_1 \,,\\[0.5cm]
l_3\,\equiv\,2\,\nt \log\rho_\e\,.
\end{array}
\eqno(5.13)
$$
Equation (5.12) is solved for $d_1$, $d_2$, and ${\w C}$ by the
Cramer's rule:
$$
\begin{array}{l}
{\w C}\,=\,\displaystyle{\frac2{{\cD}}}\,\bigl(\rho^2_\e
(l_2\,-\,l_1)\,+\,
\rho^2_\ep (l_1\,-\,l_3)\bigr)\,,\\[0.5cm]
d_1\,=\,\displaystyle{\frac1{{\cD}}}\,\bigl(a l_2\,-\,b l_1\,+\,
l_3 (b - a)\bigr)
\,,\\[0.5cm]
d_2\,=\,\displaystyle{\frac2{{\cD}}}\,\bigl(\rho^2_\e (b l_1\,-\,a
l_2)\,+\, \rho^2_\ep a l_3\bigr) \,,
\end{array}
\eqno(5.14)
$$
where ${\cD}\,\equiv \,2 a \rho^2_\ep\,+\,2 (b-a) \rho^2_\e$. The
parameters ${\w C}$ (i.e., $const$, see (5.11)), $d_1$, and $d_2$
thus obtained should be used in $f$ from (5.1).

Let us assume that $\rho_\ep$ is sufficiently small (the so-called
{\it logarithmic approximation}):
$$
1\,\ll\,\displaystyle{\kappa}\,
\ll\,\displaystyle{\frac{1}{\rho_\ep}\,\ll\,\frac{1}{\rho_\ep}
\log\frac{1}{\rho_\ep}}\,. \eqno(5.15)
$$
Then, the ratios (5.14) can be represented in the perturbative
form
$$
\begin{array}{l}
{\w C}\,\simeq\,\displaystyle{-\,\frac{4 r_4}{\kappa^2} \log
\frac{1}{\rho_\ep}}\,+\,\displaystyle{\frac{4}{\kappa^2} \Bigl({\w
r}_5\,+\,(1\,+\,q)\,r_4 \Bigr)}\,+\,
\dots\,, \\[0.5cm]
d_1\,\simeq\,\displaystyle{\frac{1}{\rho^2_\e} \Bigl(\nt \log
\frac{\rho_\e}{\rho_1}\,+\,2 \frac{r_4}{\kappa^2} \log
\frac{1}{\rho_\ep} \Bigr)}\,+\,\dots\,, \\[0.5cm]
d_2\,\simeq\,-\,\displaystyle{4 \frac{r_4}{\kappa^2} \log
\frac{1}{\rho_\ep}\,-\,2 \nt \log\frac{1}{\rho_1} }\,+\,\dots\,,
\end{array}
\eqno(5.16)
$$
where dots imply terms that tend to zero as $\rho_\ep\to 0$. In
addition, the new length parameter $\rho_1$ is intended to express
the contributions which are constant as $\rho_\ep\to 0$:
$$
2 \nt\, \log\frac {1}{\rho_1}\,\equiv\,{\w r}_1\,-\,
\displaystyle{\frac{4}{\kappa^2} \Bigl({\w r}_5\,+\,(1\,+\,q)\,r_4
\Bigr)\,, }
 \eqno(5.17)
$$
where $q$ is given by (5.11). Using (5.4) and (5.16), we obtain
sufficiently far from the core:
$$
\begin{array}{rcl}
\stackrel{(2)}{\si}_{\rho \rho}&\approx& \displaystyle{ \frac{2
\nt}{\rho^2} \left(\log\frac{\rho}{\rho_1}\,-\,
\Bigl(\frac{\rho}{\rho_\e}\Bigr)^2\log\frac{\rho_\e}{\rho_1}
\right) \,-\,4 \frac{r_4}{\kappa^2} \log\rho_\ep
\Bigl(\frac{1}{\rho^2}\,-\,\frac{1}{\rho^2_\e}\Bigr)
 }\,,\\ [0.5cm]
\stackrel{(2)}{\si}_{\phi \phi}&\approx& \displaystyle{ \frac{2
\nt}{\rho^2} \left(1\,-\,\log\frac{\rho}{\rho_1}\,-\,
\Bigl(\frac{\rho}{\rho_\e}\Bigr)^2\log\frac{\rho_\e}{\rho_1}
\right) \,+\,4 \frac{r_4}{\kappa^2} \log\rho_\ep
\Bigl(\frac{1}{\rho^2}\,+\,\frac{1}{\rho^2_\e}\Bigr)}\,.
\end{array}
\eqno(5.18)
$$

As far as the asymptotic expansions (5.18) are concerned, it is
crucial that $r_4$ can be made equal to zero. Indeed, using $r_4$
in the form (4.8) (where $p_1$ and $p_3$ from (3.32) are used, and
$p_2$ is zero), we obtain $\rho_*^2$ from the equation $r_4=0$ as
follows:
$$
\begin{array}{r}
\displaystyle{ \rho_*^2\,=\,\frac{B(\eta,
\al)}{\kappa^2}}\,,\qquad \displaystyle{ B(\eta,
\al)\,\equiv\,\frac{8(1 - \nu)(1 -
\eta)}{\al(1-\nu)\eta-2 \nu (1-\eta)}}\,,\\
[0.5cm] \displaystyle{\eta\,\equiv\, 4 (1\,+\,\nu)\,\frac{\mu^2
C_3}{1\,-\,2\mu^2 C_7}}\,, \end{array} \eqno(5.19)
$$
where $\al=1$ or $\al=3$, according to (3.32). Equations (5.19)
lead to the positivity of the coefficient-function $B(\eta, \al)$.
The validity of (3.24) holds better whenever $\kappa^{-1}$ and
$\rho_*$ are smaller. If the scale $\kappa^{-1}$ is small enough,
a bound $B_0$ can be found to restrict $B(\eta, \al)\,$:\break
$0<B(\eta, \al)\,{\stackrel{<}{_\sim}}\, B_0$. If so, $\rho_*$
from (5.19) is estimated as $\rho_* = O(\kappa^{-1})$. In terms of
$\eta$, the corresponding restriction takes the form
$$
\displaystyle{ \bigl(1+\frac{\al}{2}
\frac{1-\nu}{\nu}\bigr)^{-1}\,<\,\bigl(1+U\bigr)^{-1}
{\stackrel{<}{_\sim}}\,\eta\,<\,1\,,\qquad U\,\equiv\, \frac{\al
(1-\nu)B_0}{8 (1-\nu) + 2 \nu B_0} }\,, \eqno(5.20)
$$
where $0 <\nu\le 1/2$ for isotropic materials.

It is straightforward to rewrite (5.20) in terms of the elastic
parameters $C_3$ and $C_7$:
$$
\begin{array}{l}
a)\quad \displaystyle{ C_7\,+\,2 (1+\nu)\,C_3\,<
\frac{1}{2\mu^2}\,\,{\stackrel{<}{_\sim}}\,\,C_7\,+\,2\,(1+\nu)
\bigl(1\,+\,U\bigr)\,C_3}\,,\qquad C_3\,>\,0\,,\\
[0.5cm] b)\quad \displaystyle{ C_7\,+\,2\,(1+\nu)\bigl(1\,+\,U
\bigr)\,C_3\,\,{\stackrel{<}{_\sim}}\,\,
\frac{1}{2\mu^2}\,<\,C_7\,+\,2 (1+\nu)\,C_3}\,,\qquad C_3\,<\,0\,.
\end{array}
\eqno(5.21)
$$
Further, (5.21) can be re-expressed \cite{pfl} in terms of the
parameters $m^\prime$ and $n^\prime$ from (3.4), which can in turn
be related \cite{teod} to the third order elastic moduli of the
cubic crystals. In other words, the choice of material is
restricted by (5.21). The numerical data provided by \cite{teod}
witness that these conditions look realistic for certain
materials. However, to discuss (5.21) from the viewpoint of real
crystallography is beyond the scope of the present investigation.

Straightforward usage of (4.6), (5.1), and (5.3) leads us to the
following general expression for the stress components:
$$
\begin{array}{rcl}
\stackrel{(2)}{\si}_{\rho \rho} &=& (\stackrel{(2)}{\si}_{{\rm
far}})_{\rho \rho}\,-\,k\, \displaystyle \rho^{-1}
I^{\prime}_{{\cF}}\,,
\\ [0.5cm]
\stackrel{(2)}{\si}_{\phi \phi} &=& (\stackrel{(2)}{\si}_{{\rm
far}})_{\phi \phi}\,+\,k \displaystyle \bigl(\cN^2 I_{{\cF}}\,+\,
\rho^{-1} I^{\prime}_{{\cF}}\,-\,G \bigr) \,,
\end{array}
\eqno(5.22)
$$
where $k$ and $G$ are given by (3.21) and (4.2), correspondingly,
and $I^{\prime}_{{\cF}}$ (the prime implies the differentiation
$d/d \rho$) is expressed as follows:
$$
{\cN}^{-1}\,I^{\prime}_{{\cF}}\,=\,\displaystyle{ {\w
Y}_1({\cN}\rho)\, \int\limits^\infty_\rho J_0({\cN}t) G(\kappa
t)\,t\,dt }\, -\,\displaystyle{
J_1({\cN}\rho)\,\int\limits^\infty_\rho {\w Y}_0({\cN} t) G(\kappa
t)\,t\,dt }\,.
$$
It is assumed that ${\w Y}_1(\cdot)\equiv (\pi/2)Y_1(\cdot)$
(similar to ${\w Y}_0$ in (4.6)). Now, the subscript `far' labels
the stress components in (5.22) in order to emphasize the
non-conventional choice of $d_1$ and $d_2$, which corresponds to
(5.16), provided that the internal radius of the ring
$\cD_{[\rho_\ep, \rho_\e]}$ is small enough. We also obtain the
sum of the second order stress components:
$$
\stackrel{(2)}{\si}_{\rho \rho}\,+\,\stackrel{(2)}{\si}_{\phi
\phi}\,=\,(\stackrel{(2)}{\si}_{{\rm far}})_{\rho \rho}\,+\,
(\stackrel{(2)}{\si}_{{\rm far}})_{\phi \phi}\,+\, k \displaystyle
\bigl(\cN^2 I_{{\cF}}\,-\,G \bigr)\,.\eqno(5.23)
$$

Since $r_4$ is zero for $\rho_*^2$ from (5.19), the limit
$\rho_\ep\to 0$ is suitable in (5.16) to eliminate $\rho_\ep$.
Then, $d_1$ and $d_2$ acquire the classically known form (5.6),
though $\rho_1$ is present instead of $\rho_\c$. The limiting
value of the parameter $const$ (see (4.2)) is also expressed with
the help of (5.11) and (5.16), provided that $\rho_\ep$ decreases
to zero. The large-distance stresses (5.18) take at $r_4=0$ the
classical form (5.7). Note that the axis $Oz$ is not captured by
the present approach, as far as the limiting transition is
allowed. This is in agreement with the regularization of the
density profile at $\rho=0$ in Section 3. The components
$(\stackrel{(2)}{\si}_{{\rm far}})_{\rho \rho}$ and
$(\stackrel{(2)}{\si}_{{\rm far}})_{\phi \phi}$ reduce, after the
shrinking of the internal boundary of the ring, to (5.18) (with
$r_4$ equated to zero) and should be distinguished from the
analogous ones given by (5.7). Eventually, there is no cut-off
around the dislocation's core for the second order stresses
expressed by (5.22), provided that the numerical parameters are
substituted appropriately.

\subsection{The component $\si_{z z}$}

Let us consider the stress component $\stackrel{(2)}{\si}_{z
z}=p$. Classically, the second equation in (3.20) implies the
following representation \cite{pfl}:
$$
\displaystyle{  \stackrel{(2)}{\si}_{z z}\,=\,\nu
\,\bigl(\stackrel{(2)}{\si}_{\rho \rho}\,+\,
\stackrel{(2)}{\si}_{\phi \phi}\bigr)\,-\, 2 \mu (1+\nu)
\bigl(\Psi^{{\rm bg}}_{z z}\,+\,Q^\prime\bigr)\,, } \eqno(5.24)
$$
where $f_{{\rm bg}}$ from (5.1.2) is used for the stresses on the
right-hand side (the respective subscript is omitted for brevity),
and $\stackrel{(1)}{\phi}_{{\rm bg}}= (-b/2\pi) \log\rho$ is
substituted to express $\Psi^{{\rm bg}}_{z z}$. As far as the
equation (5.24) is concerned, it is assumed that the argument
$\rho$ varies within the segment $[\rho_\c, \rho_\e]$, and
$Q^\prime$ is determined from the requirement that
$\stackrel{(2)}{\si}_{z z}$ averaged over $\cD_{[\rho_\c,
\rho_\e]}$ is zero (the so-called `mean stress theorem'; see
\cite{teod}). Then, the representation (5.24) reads for $\rho_\e
\gg \rho_\c$:
$$
\displaystyle{ \stackrel{(2)}{\si}_{z z}\,\approx\,2\,\biggl[ \nu
\nt\,+\,\Bigl(\frac{b}{2\pi}\Bigr)^2 \,\mu^3 (1+\nu)
C_3\biggr]\,\biggl[\frac1{\rho^2}\,-\,
\frac2{\rho^2_\e}\,\log\frac{\rho_\e}{\rho_\c}\biggr]\,. }
\eqno(5.25)
$$

Let us turn to the stress component $\stackrel{(2)}{\si}_{z z}$
from (3.26.2) of the modified defect:
$$
\displaystyle{
\stackrel{(2)}{\si}_{z z}\,=\,\nu\,
\bigl(\stackrel{(2)}{\si}_{\rho \rho}\,+\,
\stackrel{(2)}{\si}_{\phi \phi}\bigr)\,-\,
            2\mu\,(1+\nu)
(\Psi_{z z}\,+\,{\w g})\,+\,\kappa^2\,(1+\nu)\,(f-f_{{\rm bg}})
}\,. \eqno(5.26)
$$
Now $\rho$ varies within $[\rho_\ep, \rho_\e]$. From the previous
considerations it is clear that $\stackrel{(2)}{\si}_{\rho \rho} +
\stackrel{(2)}{\si}_{\phi \phi}$ and $\Psi_{z z}$, taken at
$\rho=\rho_\ep$, decrease as $\rho_\ep\to 0$. Nevertheless,
$\stackrel{(2)} {\si}_{z z}$ behaves, sufficiently close to $Oz$,
somewhat artificially. The reason is that the logarithmic terms
dominate in $f-f_{{\rm bg}}$ at $\rho$ small enough. However, this
is not an obstacle to view the rings $\cD_{[\rho_\ep, \rho_\e]}$,
as well as (in the limiting sense) the punctured disc, as the
domains of definition of the stress distributions. Indeed,
$\stackrel{(2)}{\si}_{z z}$ from (5.26), averaged over
$\cD_{[\rho_\ep, \rho_\e]}$, is finite, provided that $\rho_\ep$
goes to zero.

Let us obtain $\stackrel{(2)}{\si}_{z z}$ from (5.26) at
sufficiently large distances, where $f-f_{{\rm bg}}$ ($= k
I_{{\cF}}$) is exponentially small. Using (5.18) at $r_4=0$, we
obtain
$$
\stackrel{(2)}{\si}_{z z}\,\approx\, 2 \nu
\nt\,\biggl[\displaystyle{ \frac{1}{\rho^2}\,-\,
\frac{2}{\rho^2_\e}\log\frac{\rho_\e}{\rho_1} \biggr]\,-\,2\mu
(1+\nu) \biggl[Q^{\prime}\,-\,\Bigl(\frac{b}{2\pi}\Bigr)^2\,
\frac{c}{\rho^2}\biggr]}\,. \eqno(5.27)
$$
We determine $Q^{\prime}$ approximately:
$$
Q^{\prime}\,\approx\,\displaystyle{ \Bigl(\frac{b}{2
\pi}\Bigr)^2\,2 c\,\frac{\log (\rho_\e /\rho_2)}{\rho^2_\e}}\,,
\eqno(5.28)
$$
where another length $\rho_2$ is introduced, namely,
$$
\begin{array}{rcl}
\displaystyle{ \Bigl(\frac{b}{2 \pi}\Bigr)^2\,2 c\,\log\frac{ 1}
{\rho_2}}&\equiv & \displaystyle{ \Bigl(\frac{b}{4
\pi}\Bigr)^2\,-\, 2 \,\int\limits_0^{1}\Psi_{z z}\,\rho\,d \rho}\\
[0.5cm] &-&\,\displaystyle{ 2
\,\int\limits_1^{\infty}\biggl[\Psi_{z
z}+\Bigl(\frac{b}{2\pi}\Bigr)^2\,\frac{c}{\rho^2}\biggr]\,\rho\,d
\rho\,+\,2 \kappa^2\,\frac{1+\nu}{1-\nu} \,\int\limits_0^{\infty}
I_{{\cF}}\,\rho\,d \rho}\,,
\end{array}
\eqno(5.29)
$$
provided that $\rho_\ep\to 0$, and the upper integration bound
$\rho_\e$ is replaced approximately by infinity.

Therefore, (5.27) depends on two scales, which are, in principle,
different in our consideration: $\rho_1$ from (5.17) and $\rho_2$
from (5.29). To re-write (5.27) more conventionally, it is
appropriate to introduce another length, say $\rho_\m$, as
follows:
$$
\begin{array}{rcl}
\displaystyle{ \biggl[ \nu \nt\,+\, \Bigl(\frac{b}{2\pi}\Bigr)^2
\mu (1+\nu) c \biggr]}\log \displaystyle{\frac{1}{\rho_\m}} & &
\\[0.5cm]
\equiv \displaystyle{\nu \nt\,\log\frac{1}{\rho_1}} &+&
\displaystyle{\Bigl(\frac{b}{2 \pi}\Bigr)^2\,\mu (1+\nu) c
\,\log\frac{1}{\rho_2}}\,.
\end{array}
\eqno(5.30)
$$
Substituting $Q^{\prime}$ from (5.28) into (5.27) and using
(5.30), we obtain
$$
\displaystyle{ \stackrel{(2)}{\si}_{z z}\,\approx\, 2\,\biggl[ \nu
\nt\,+\,\Bigl(\frac{b}{2\pi}\Bigr)^2 \mu (1+\nu) c
\biggr]\,\biggl[\frac1{\rho^2}\,-\,
\frac2{\rho^2_\e}\,\log\frac{\rho_\e}{\rho_\m}\biggr]\,. }
\eqno(5.31)
$$
Equation (5.31) looks similarly to (5.25), except that now
$\rho_\m$ is present instead of $\rho_\c$.

The classical answers for $d_1$ and $d_2$ are given at
$\rho_\e\gg\rho_\c$ by (5.6). Since $d_1$ and $d_2$ from (5.16)
include the constant contributions, the agreement between (5.6)
and (5.16) is due to the newly defined length $\rho_1$ from
(5.17). The mean stress theorem implies that $\rho_2$ coincides,
classically, with the lower integration boundary $\rho_\c$,
provided that $\rho_\c$ remains unspecified. Instead, $Q^\prime$
in (5.28) is concerned with another length $\rho_2$ from (5.29).
This is because $\Psi_{z z}$ is unconventional inside the core. In
the present approach, two different lengths arise. The definition
(5.30) is only to obtain the classically-looking representation
(5.31).

We use ${\w g}$ (3.25), where $\cC=Q^\prime$ with $Q^\prime$ given
by (5.28), and we rewrite the stress $\stackrel{(2)}{\si}_{z z}$
expressed by (5.26):
$$
\begin{array}{rcl}
\stackrel{(2)}{\si}_{z z}\, &= & \displaystyle{
\nu\,\bigl(\stackrel{(2)}{\si}_{\rho
\rho}\,+\,\stackrel{(2)}{\si}_{\phi \phi}\bigr)\,-\,2\mu\,(1+\nu)
\biggl[\Psi_{z z}\,-\,\frac2{\rho^2_\e}
\,\int\limits_0^{\rho_\e}\Psi_{z z}\,\rho\,d \rho\biggr]} \\[0.5cm]
&-& \displaystyle{ 2\mu\,(1+\nu)\Bigl(\frac{b}{4
\pi}\Bigr)^2\,\biggl[\frac{1}{\rho^2_\e}\,+\, \frac{2}{\rho^2_*}\,
\Bigr(\frac{\rho^2}{\rho^2_*}\,-\,1\Bigl)\,
h_{[0,\,\rho_*]}(\rho)\biggr]
} \\[0.5cm]
&+& \displaystyle{2 \kappa^2\,\frac{\mu (1+\nu)^2}{1 - \nu}
\,\biggl[ I_{{\cF}}\,-\,\frac2{\rho^2_\e}
\,\int\limits_0^{\rho_\e} I_{{\cF}}\,\rho\,d \rho\biggr] }\,.
\end{array}
\eqno(5.32)
$$
Moreover, using (5.23) we can further rewrite (5.32) for
sufficiently large $\rho_\e$ as follows:
$$
\begin{array}{rcl}
\stackrel{(2)}{\si}_{z z}\, &\approx & \displaystyle{ \nu\,\bigl(
(\stackrel{(2)}{\si}_{{\rm far}})_{\rho \rho}\,+\,
(\stackrel{(2)}{\si}_{{\rm far}})_{\phi \phi} \bigr)\,-\,2
\mu\,(1+\nu) \Psi_{z z} }
\\[0.5cm]
&-& \displaystyle{ 2\mu\,(1+\nu)\Bigl(\frac{b}{2
\pi}\Bigr)^2\,\left[\frac{2
c}{\rho^2_\e}\,\log\frac{\rho_\e}{\rho_2}\,+\, \frac{1}{2
\rho^2_*}\, \Bigl(\frac{\rho^2}{\rho^2_*}\,-\,1\Bigr)\,
h_{[0,\,\rho_*]}(\rho)\right] }
\\[0.5cm]
&+& k\,\displaystyle \bigl(\cN^2 I_{{\cF}}\,-\,\nu\,G \bigr) \,,
\end{array}
\eqno(5.33)
$$
where $k$, $G$, $I_{{\cF}}$, and $\rho_*^2$ are given by (3.21),
(4.2), (4.6), and (5.19), respectively. Expression for $\Psi_{z
z}$ from (3.18) in the polar coordinates takes the form: $\Psi_{z
z} = -c(\cd_{\rho} \stackrel{(1)}{\phi})^2$, where
$\stackrel{(1)}{\phi}$ is given by (3.13). The behaviour of
$\Psi_{z z}$ for sufficiently large distances is similar to that
of $\Psi^{{\rm bg}}_{z z}$ although is different within the core.

Let us recall that a stationary Schr\"odinger equation obtained in
the effective mass approximation \cite{bar} has been used for the
conduction electrons in the dislocated \cite{osip} or disclinated
crystals \cite{osip1}, \cite{osip3}. The influence of the defects
has been accounted for via the deformation potential given by the
trace of the strain tensor (i.e., by the dilatation). The gauge
and/or geometric approaches to various effects due to the
dislocations (due to the torsion) attract considerable attention:
\cite{kaw}--\cite{klein1}, \cite{fur2}, and
\cite{azev1}\footnote{For more refs. on other types of defects,
see \cite{mok}.}.

The dilatation is non-trivial for the screw dislocation just
because of the strains of the second order. Sufficiently far from
the core, the dilatation depends on $1/\rho^2$ just by means of
$m^\prime$ and $n^\prime$ from (3.4) (see \cite{pfl}). The results
obtained above can be applied to the Schr\"odinger equation, as is
discussed in \cite{osip}. Indeed, the solution \cite{osip} is
valid only outside the core, and, say, a traction-free condition
is still required. In turn, a possible influence of the
dilatation, obtainable by the present approach, on a local shape
of the wave function can be investigated in the vicinity of the
core region. The corresponding results should be comparable with
appropriate lattice-based investigations. Note that the
disclination core's radii have been used in \cite{osip1} and
\cite{fur4} for description of the corresponding electron
localization.

\subsection{Remarks on the logarithmic approximation}

The asymptotic relations (5.16) give an indication of the fact
that the ratios (5.14) should be re-arranged perturbatively for
$\log(1/\rho_\ep)$ and $\rho_\e$ sufficiently large. It is
instructive to re-derive the leading terms in (5.16) by means of a
more straightforward use of the logarithmic approximation at
$r_4=0$. Indeed, let us reconsider (5.12), provided that the
parameters (5.13) are taken in the principal order as follows:
$$
\begin{array}{ll}
\displaystyle{ a \,\approx\,- 1}\,, & \qquad\displaystyle{ b
\,\approx\,- 1\,-\,\frac{(\kappa\rho_\ep)^2}2 \,\log\rho_\ep
}\,,\\[0.5cm]
l_1\,\approx\,-\,{\w r}_1\,, & \qquad l_2\,\approx\,-\,{\w
r}_1\,-\,2\,{\overset \o{r}}_5 \rho_\ep^2 \log\rho_\ep
 \,.
 \end{array}
 \eqno(5.34)
$$
Here, $r_4=0$ in ${\w r}_1$, and the special notation ${\overset
\o{r}}_5$ is chosen for ${\w r}_5$ at $r_4=0$. Besides, $l_3$ is
kept like in (5.13). We replace in (5.12) the entry equal to
$2\rho_\ep^2$ by zero. Then, in agreement with (5.16), the
solution reads:
$$
\displaystyle{ {\w C}\,=\,\frac{4\,{\overset
\o{r}}_5}{\kappa^2}\,, \qquad d_1\,=\,\frac{\nt}{\rho^2_\e}\,
\log\frac{\rho_\e}{\rho_1}\,,\qquad\, d_2\,=\,2 \nt\,\log\rho_1
}\,, \eqno(5.35)
$$
where $\rho_1$ agrees with (5.17).

Moreover, one can estimate $\rho_1$, provided that $\log\kappa$ is
large enough. With regard to the definition by means of the
equation (5.10), we extract the dependence of ${\w r}_1$ and ${\w
r}_6$ on the large logarithm as follows ($r_4$ is zero):
$$
{\w r}_1\,=\,\frac{q_1}{\kappa^2}\,\log\kappa\,+\,{\overset
\o{r}}_1\,,\qquad {\w r}_6\,=\,{\overset
\o{r}}_5\,\log\kappa\,+\,{\overset \o{r}}_6\,. \eqno(5.36)
$$
The terms ${\overset \o{r}}_1$ and ${\overset \o{r}}_6$ in (5.36)
are simply intended to denote the corresponding remnants, which
are finite, provided that $\log\kappa$ is growing. We can use
(5.36) in (5.13) and again directly solve (5.12). Then, $d_1$ and
$d_2$ appear just in the form (5.35) with $\rho_1=B_1/\kappa$,
where
$$
2\nt\log B_1\,\equiv\, -\,{\overset
\o{r}}_1\,+\,\frac{4}{\kappa^2}\,{\overset \o{r}}_5\,. \eqno(5.37)
$$
It is seen that (5.37) agrees with (5.17). The radius $\rho_2$ can
also be estimated. From (5.29) we obtain $\rho_2 =B_2/\kappa$,
where
$$
2 c\,\log B_2\,\equiv\,-\,\frac{1}{4}\, -\,2
c\,\int\limits^1_0(1\,-\,s\,K_1(s))^2\frac{ds}{s}\,, \eqno(5.38)
$$
and $c$ is given by (3.30). Therefore, the logarithmic
approximation allows us to determine the leading behaviour of ${\w
C}$, $d_1$, and $d_2$, as well as to estimate $\rho_1$ and
$\rho_2$. Note in passing that a certain influence of the
regularization in (4.2) on $B_1$ from (5.37) and $B_2$ from (5.38)
should be accounted for, though elsewhere (since specific
numerical estimates based on the crystallographic data are
required).

The result (5.35) should be commented as follows. Let us introduce
a sufficiently small positive $\in\,$ by the requirement: absolute
values of the stresses, which are smaller than $\in$ (say,
$|\stackrel{(2)}{\si}_{i j}| \le\,\displaystyle{\frac{\in}2}$),
should be treated, for engineering, as indistinguishable from
zero. Then, the shrinking of the boundary at $\rho = \rho_\ep$ can
be ceased just for those boundary values $\stackrel{(2)}{\si}_{i
j}\Big|_{\rho=\rho_\ep}$, which respect this estimate. At the same
time, (5.15) holds, and thus (5.35), (5.37), and (5.38) are valid.
The large distance behavior looks conventional though includes now
$\rho_1$ and $\rho_2$, which are dictated by a choice of material.
Provided $\rho_\ep$ is thus fixed, the stress
$\stackrel{(2)}{\si}_{z z}$ remains bounded outside an
infinitesimally thin tube around $Oz$. It is essential that under
the prescribed accuracy $\in$, a distant ``observer'' outside the
core is dealt only with the physical stresses improved by means of
$\rho_1$ and $\rho_2$. These stresses are insensitive to
particular $\rho_\ep$'s.

Let us turn again to our geometric interpretation. Let us imagine
for a moment that the differential-geometric torsion is taken into
account in order to make (3.16) conventionally looking. Then,
$\Dl\stackrel{(1)}{\phi}_\cT$ comes to play instead of
$\Dl\stackrel{(1)}{\phi}$. Since a specific shape of appropriately
localized density $\cT$ is unclear, (3.24) can be applied, as a
simplification, just to $\Dl\stackrel{(1)}{\phi}_\cT$. As a
result, the driving source $\cR$ should be changed. The local
properties of this new $\cR$ can be estimated. The modifications
are expectable for $p_4$, $p_5$, and $p_6$ from (3.31). However,
these changes do not influence the coefficients we consider in
(4.7). A change of shapes of the profiles of the stress components
is expectable in the middle of the core, but this is beyond our
scope. It is important that $\rho_1$ in (5.17) and $\rho_*$ in
(5.19) should not be influenced when $\Dl\stackrel{(1)}{\phi}_\cT$
is subject to (3.24). Only $\rho_2$ from (5.29) should be
replaced, since it is defined ``globally''. Thus, the present
picture is almost the same as that with the torsion admitted but
handled approximately. To avoid a specification of $\cT$, it is
appropriate to keep our teleparallel interpretation.

\section{ Discussion}

The stress potential of the non-singular screw dislocation is
found in the quadratic approximation. It is given by the sum of
the conventional part and the gauge contribution. The general
expression (5.22) for the stress components of second order
$\stackrel{(2)}{\si}_{\rho \rho}$ and $\stackrel{(2)}{\si}_{\phi
\phi}$ is also obtained by means of appropriate differentiations
of the stress potential. In addition, the general expression for
the stress $\stackrel{(2)}{\si}_{z z}$ is elaborated as well: see,
for instance, (5.32) or (5.33). The main attention is paid to the
asymptotic properties of the stresses in question. To ensure their
self-consistency, the arbitrariness of certain constants in the
general solution for the stress potential is used.

Sufficiently far from the core, the second order stresses
obtained, $\stackrel{(2)}{\si}_{\rho \rho}$,
$\stackrel{(2)}{\si}_{\phi \phi}$, and $\stackrel{(2)}{\si}_{z z}$
are in agreement with \cite{pfl}, although now they include the
new lengths $\rho_1$ and $\rho_2$. The non-conventional part of
the total stress potential is responsible for the localized
contribution into the full stress distribution. As a result, the
modification of the short-distance behaviour is as follows:
$\stackrel{(2)}{\si}_{\rho \rho}$ and $\stackrel{(2)}{\si}_{\phi
\phi}$ tend to zero, while $\stackrel{(2)}{\si}_{z z}$ grows
logarithmically (although $\stackrel{(2)}{\si}_{z z}$ is
integrable, as required in \cite{pfl}) in the close vicinity of
$Oz$. There is no artificial cut-off at an internal boundary. The
``exterior'' (with respect to the core region) solution for the
second-order stresses plays a central role in the picture
presented. This is because of its improvement due to the
self-consistent arising of $\rho_1$ and $\rho_2$. The general
solution is characterized by one more length $\rho_*$ which is a
half-width of the effective defect's density profile.

The lengths $\rho_1$, $\rho_2$, and $\rho_*$ are expressed through
the elastic moduli of second and third order. The appearance of
several lengths correlates with the absence of a sharp boundary
around the core in the first order pictures in \cite{ed3},
\cite{mal}, \cite{laz2}, and \cite{laz4}. In other words, a
concept of `transition shell' looks helpful for characterization
of the solution obtained. This shell would separate two domains: a
compact region under the shell, where the strains are rather
finite, and an outer bulk. The solution obtained should be less
reliable under the shell (where the effects of discreteness are
comparable with those of strong elastic non-linearity; see
\cite{hir} and \cite{teod}). Within the transition shell, the
quadratic corrections both of the conventional and of the gauge
origin (i.e., due to $I_{{\cF}}$ in the stress potential) should
be important. Outside the shell (i.e., in the bulk), the
conventional terms become more valuable, since the gauge
contributions are strongly decaying. The radii obtained $\rho_1$,
$\rho_2$, and $\rho_*$ should characterize the location and the
extent of the transition shell. More generally, they seem to
display the dislocation cores as radially-layered regions.

The radii $\rho_*$, $\rho_1$, and $\rho_2$ include the basic
length $1/\kappa$ ($\,\equiv {\sqrt{ \ell/\mu}}\,$), which can be
adjusted to the interatomic scale even in the first order: see,
for instance, \cite{ed3} and \cite{laz2}. Analogous
identifications can also be found in \cite{cem}, \cite{g22}, and
\cite{g4}. Here $\ell$ is the gauge-material parameter (see
(2.10)) and $\mu$ is the shear modulus. However, the corresponding
coefficients $B(\eta, \al)$, $B_1$, and $B_2$ given by (5.19),
(5.37), and (5.38), respectively, are strictly influenced by the
crystallographic moduli of second and third order. Mutual
comparisons of the coefficients and more specific statements on
the properties of the transition shell should be done elsewhere.
The numerically obtainable values for $\rho_1$, $\rho_2$, and
$\rho_*$ can be compared, in principle, with appropriate scales
provided by the semi-discrete models of the cores (see refs. in
\cite{hir}, \cite{teod}, and \cite{gair}). On the other hand,
attempts at matching the radii in accordance with the experimental
observations may result in independent estimate for $1/\kappa$.
Such estimate can, in turn, be compared with $1/\kappa$ obtainable
by means of the first-order considerations.

As far as the numerical estimates are concerned, the isotropy
requirement should be taken into account, since the number of the
elastic constants of third order for the cubic crystals is six
(see \cite{teod} and \cite{gair}). Moreover, strong couplings
could modify the elastic parameters within the core in comparison
with those outside it (see \cite{kun}). Note that \cite{marad}
also provides the radius of the screw dislocation as a function of
the elastic constants (of second order though) and of the lattice
spacings of the crystal.

Strains and displacements within the transition shell can, in
principle, be elaborated with the help of the present approach.
For instance, the displacements of the first order in the core
region are discussed in \cite{ed3} and \cite{laz2}. The next order
corrections to those displacements can be obtained within the
shell. Here, the technical issue \cite{ed4} could be helpful. Thus
corrected displacements should be comparable with the
semi-discrete calculations subject to the so-called flexible
boundary conditions (see \cite{sin} and \cite{geh}). Direct
observations of the dislocation cores by means of the high
resolution electron microscopy can also be used for comparison
(for more refs. on the observation and modelling of the
dislocation cores at atomic level in semiconductors, for instance,
see \cite{xu}). The solution obtained can also be tested for the
problems where just the second-order elasticity is relevant, e.g.,
for the elastic waves or the electron (see Sec. 5.2) scattering,
etc. (see \cite{teod} and \cite{haif}).

The continuation obtained for the stresses is just due to the
gauge equation considered as the incompatibility law. In turn, our
constitutive equations imply that the elastic energy is written up
to the third order. But the quadratic constitutive law is only an
approximation for essentially non-linear situation. Clearly, the
present picture does not pretend to simulate the stress
distribution in the middle of the core, where both the atomistic
structure and the finite elasticity are crucial (see \cite{pet}
where a non-linear elasticity approach with flexible boundary is
discussed as a way for improving the semi-discrete techniques).
Nevertheless, the stress distribution provided by the present
model of the screw dislocation is self-contained outside a certain
shell, and it seemingly remains satisfactory within this shell as
well.

Our investigation demonstrates that the Hilbert--Einsteinian gauge
approach is flexible enough, since it allows, in two orders, the
self-consistent description for the non-singular screw
dislocation. Certain features of an extended picture could be
discovered in the present model. It can be used as a base for
further, more involved, treatments. Application to the edge
dislocation would be also desirable.

\section*{Acknowledgement}

The author is grateful to M. Lazar for interesting and useful
correspondence on the defects and for comments on \cite{ml}, as
well as to N. M. Bogoliubov, M. Yu. Gutkin, A. L. Kolesnikova, and
A. E. Romanov for discussing the paper during its
preparation\footnote{The author is grateful to M. Lazar for
calling attention to \cite{sahoo1} where the modified screw
dislocation has also been obtained for the gauge Lagrangian of the
Maxwell type.}. The research described has been supported in part
by the RFBR ($\#$ 01--01--01045 and $\#$ 04--01--00825) and by the
Russian Academy of Sciences program ``Mathematical Methods in
Non-Linear Dynamics''.

\section*{Appendix A}

Appendices A and B contain intermediate results, which are helpful
for expanding the stress potential $f$ in (5.1) (see also
\cite{ml}). Starting with the series expansion (3.31) and using
the Bessel functions in the series form \cite{wat}, we go over to
the corresponding series $\cF$ from (4.7).

First, let us estimate $G(s)$ from (4.2). To begin with, we expand
the products \break $t^{-1} K_0(t) {\cR}(t)$ and $t^{-1} I_0(t)
{\cR}(t)$ for a small argument $t\ll 1$:
$$
\begin{array}{rcl}
t^{-1} I_0(t) {\cR}(t) &\simeq&
\displaystyle{
p_1\,t^{-3}\,+\,p_2\,t^{-1}
\log t\,+\,\left(\frac{p_1}4\,+\,p_3\right) t^{-1}
}          \\[0.5cm]
&+&
\displaystyle{
\left( p_4 \log^2 t\,+\,\Bigl(\frac{p_2}4\,+\,
    p_5\Bigr)\,\log t\,+\,{\h k} \right) t\,+\,\dots
}\,;
\end{array}
\eqno({\rm A}1)
$$
$$
\begin{array}{rcl}
t^{-1} K_0(t) {\cR}(t) &\simeq&
\displaystyle{
- p_1\,\log\Bigl(\frac{\ga}2\,t\Bigr)\,t^{-3}\,-\,
\left( p_2 \log^2 t\,+\,k_1\,\log t\,+\,k_2 \right) t^{-1}
}          \\[0.5cm]
&-&
\displaystyle{
\left( p_4 \log^3 t\,+\,k_3 \log^2 t\,+\,k_4\,\log t\,+\,
k_5 \right) t\,+\,\dots
}\,,
\end{array}
\eqno({\rm A}2)
$$
where $p_1, p_2, \dots, p_5$ are given by (3.32), and the
coefficients $k_1, k_2$, and $k_3$ are expressed as follows:
$$
\begin{array}{l}
\displaystyle{
k_1\,=\,\frac{p_1}4\,+\,\log\Bigl(\frac\ga{2}\Bigr)\,p_2\,+\,p_3
}\,,\\[0.5cm]
\displaystyle{
k_2\,=\,-\frac{p_1}4\,+\,\log\Bigl(\frac\ga{2}\Bigr)\,
\left( \frac{p_1}4\,+\,p_3\right)
}\,,\\[0.5cm]
\displaystyle{
k_3\,=\,\frac{p_2}4\,+\,\log\Bigl(\frac\ga{2}\Bigr)\,p_4\,+\,p_5 }
\end{array}
\eqno({\rm A}3)
$$
($\gamma$ is the Euler constant). The parameters $k_4$, $k_5$, and
${\h k}$ are not used in the present paper and can be found in
\cite{ml}.

Using (A1) and (A2), we obtain the following estimates for $s \ll
1$:
$$
\begin{array}{rcl}
\displaystyle{ K_0 (s) \biggl( B(\epsilon) -
\int\limits_\epsilon^s I_0(t) {\cR}(t)\,\frac{dt}{t}\biggr)\Bigg
|_{\epsilon \to 0}} &\simeq & \displaystyle{
s^{-2}\,\left({\cK}_1\,\log s\,+\,{\cK}_2\right)
}             \\[0.5cm]
&+& \displaystyle{\sum_{i=0}^3 {\cK}_{6-i}\,\log^i
s\,+\,s^2\,\sum_{i=0}^3 {\cK}_{10-i}\,\log^i s\,+\,\dots \,, }
\end{array}
\eqno({\rm A}4)
$$
where
$$
\begin{array}{l}
\displaystyle{ {\cK}_1\,=\,\frac{-p_1}2\,, \qquad
{\cK}_2\,=\,-\log\Bigl(\frac\ga{2}\Bigr)\,\frac{p_1}2\,, \qquad
{\cK}_3\,=\,\frac{p_2}2\,,\qquad
{\cK}_4\,=\,\frac{p_1}4\,+\,\log\Bigl(\frac\ga{2}\Bigr)\,
\frac{p_2}2\,+\,p_3\,,
}\\[0.5cm]
\displaystyle{
 {\cK}_5\,=\,-\,{\rm const}\,+\,\Bigl(\frac{3}2 +
\log\frac\ga{2}\Bigr)\,
\frac{p_1}4\,+\,\log\Bigl(\frac\ga{2}\Bigr)\,p_3 \,,
}\\[0.5cm]
\displaystyle{ {\cK}_6\,=\,-\,{\rm const}\times
\log\frac\ga{2}\,+\, \Bigl(1\,+\,3
\log\frac\ga{2}\Bigr)\,\frac{p_1}8\,, \qquad \qquad \qquad
{\cK}_7\,=\,\frac{p_2}8\,+\,\frac{p_4}2\,,
}\\[0.5cm]
\displaystyle{
{\cK}_8\,=\,\frac{p_1}{16}\,+\,\log\Bigl(\frac\ga{2}\Bigr)\,
\frac{p_2}{8}\,+\,\frac{p_3}{4}\,-\,
\Bigl(1\,-\,\log\frac\ga{2}\Bigr)\,\frac{p_4}2\,+\,\frac{p_5}2\,.
}
\end{array}
\eqno({\rm A}5)
$$
Further, we get
$$
\begin{array}{rcl}
\displaystyle{ I_0 (s)\,\int\limits_s^\infty K_0(t)
{\cR}(t)\,\frac{dt}{t}} & \simeq & \displaystyle{
s^{-2}\,\left({\cI}_1\,\log s\,+\,{\cI}_2\right)
}             \\[0.5cm]
&+& \displaystyle{\sum_{i=0}^3 {\cI}_{6-i}\,\log^i
s\,+\,s^2\,\sum_{i=0}^3 {\cI}_{10-i}\,\log^i s\,+\,\dots \,, }
\end{array}
\eqno({\rm A}6)
$$
where
$$
\begin{array}{l}
\displaystyle{ {\cI}_1\,=\,\frac{- p_1}{2}\,,\qquad {\cI}_2\,=\,-
\Bigl(1 + 2 \log\frac\ga{2}\Bigr)\,\frac{p_1}4 \,,\qquad
{\cI}_3\,=\,\frac{p_2}3\,,\qquad {\cI}_4\,=\,\frac{k_1}2 \,,
}\\[0.5cm]
\displaystyle{ {\cI}_5\,=\,\frac{-p_1}8\,+\,k_2\,,\qquad \qquad
\quad {\cI}_6\,=\,{\cI}_K\,+\,\Bigl(1 + 2 \log\frac\ga{2}\Bigr)\,
\frac{3 p_1}{16}\,,
}\\[0.5cm]
\displaystyle{ {\cI}_7\,=\,\frac{p_2}{12}\,+\,\frac{p_4}2
\,,\qquad \qquad \qquad {\cI}_8\,=\,\frac{k_1}8\,-\,\frac{3
p_4}4\,+\,\frac{k_3}2 \,, }
\end{array}
\eqno({\rm A}7)
$$
and $k_1$, $k_2$, and $k_3$ are given by (A3). The numerical
constant ${\cI}_K$ in ${\cI}_6$ from (A7) implies the regularized
value of the integral:
$$
\begin{array}{rcl}
{\cI}_K &\equiv& \displaystyle{ \int\limits_1^\infty K_0(t)
{\cR}(t)\,\frac{dt}t\,+\, \int\limits_0^1 \Bigl[ K_0(t)
{\cR}(t)\,+\,p_1\, \log\Bigl(\frac{\ga}2\,t\Bigr)\,t^{-2} \Bigr.}
\\[0.5cm]
&+& \displaystyle{ \Bigl.k_2\,+\,k_1\,\log t\,+\,p_2\,\log^2
t\Bigr]\frac{dt}t\,. }
\end{array}
\eqno({\rm A}8)
$$

Finally, the series (4.3) is obtained by subtracting the series
(A6) from (A4). The resulting coefficients (4.4) are calculated by
means of (A5) and (A7): $ q_i\,=\,{\cK}_{i+1}\,-\,{\cI}_{i+1}$ for
$i = 1, 2, \dots\,, 7$. Since ${\cK}_1-{\cI}_1=0$, the
contribution $\propto s^{-2} \log s$ is absent in (4.3).

\section*{Appendix B}

Now let us estimate $I_{{\cF}}(\rho)$ from (4.6). First, we expand
the products $t\,J_0({\cN}t) G(\kappa t)$ and \break $t\,{\w
Y}_0({\cN}t) G(\kappa t)$ for a small argument $t\ll 1$:
$$
\begin{array}{rcl}
t\,J_0({\cN}t)\,G(\kappa t)&\simeq & \displaystyle{
\frac{q_1}{\kappa^2}\,t^{-1}\,+\,t\,\Bigl(q_2
\log^3\bigl({\cN}t\bigr)
}\\[0.5cm]
&+&\!\!\!\displaystyle{ \sum_{i=0}^2
n_{3-i}\,\log^i\bigl({\cN}t\bigr)\Bigr)\,+\,t^3\,\sum_{i=0}^3
n_{7-i}\,\log^i\bigl({\cN}t\bigr)\,+\,\dots\,, }
\end{array}
\eqno({\rm B}1)
$$
\vskip 0.4cm
$$
\begin{array}{rcl}
t\,{\w Y}_0({\cN}t)\,G(\kappa t) &\simeq & \displaystyle{
\frac{q_1}{\kappa^2}\,t^{-1}\,\log\Bigl(\frac{\ga}2 {\cN}
t\Bigr)\,+\,t\,\Bigl(q_2\,\log^4\bigl({\cN}t\bigr)
}\\[0.5cm]
&+&\!\!\!\displaystyle{\sum_{i=0}^3
m_{4-i}\,\log^i\bigl({\cN}t\bigr)\Bigr)\,+\,t^3\,\sum_{i=0}^4
m_{9-i}\,\log^i\bigl({\cN}t\bigr)\,+\,\dots\,, }
\end{array}
\eqno({\rm B}2)
$$
where the parameters $n_1, n_2, \dots , n_5$ are expressed by
means of the previously found coefficients (4.4):
$$
\begin{array}{l}
\displaystyle{
n_1\,=\,3\,\log\Bigl(\frac{\kappa}{{\cN}}\Bigr)\,q_2\,+\,q_3\,,
}\\[0.5cm]
\displaystyle{
n_2\,=\,3\,\log^2\Bigl(\frac{\kappa}{{\cN}}\Bigr)\,q_2
\,+\,2\,\log\Bigl(\frac{\kappa}{{\cN}}\Bigr)q_3\,+\,q_4\,,
}   \\[0.5cm]
\displaystyle{
n_3\,=\,-\,\frac{{\cN}^2}{\kappa^2}\,\frac{q_1}4\,+\,
\log^3\Bigl(\frac{\kappa}{{\cN}}\Bigr)\,q_2
\,+\,\log^2\Bigl(\frac{\kappa}{{\cN}}\Bigr)q_3
\,+\,\log\Bigl(\frac{\kappa}{{\cN}}\Bigr)q_4\,+\,q_5\,,
}\\[0.5cm]
\displaystyle{
n_4\,=\,-\,{\cN}^2\,\frac{q_2}4\,+\,\kappa^2\,q_6\,,
}            \\[0.5cm]
\displaystyle{
n_5\,=\,-\,{\cN}^2\,\frac{q_3}4\,+\,\kappa^2\,q_7\,+\,
3\,\log\Bigl(\frac{\kappa}{{\cN}}\Bigr)\,n_4\,.
}            \\[0.5cm]
\end{array}
\eqno({\rm B}3)
$$
The coefficients $m_1, m_2, \dots, m_6$ are expressed by means of
$n_1, n_2, \dots, n_5$ and $q_1$ and $q_2$:
$$
\begin{array}{l}
\displaystyle{
m_1\,=\,\log\Bigl(\frac{\ga}2\Bigr)\,q_2\,+\,n_1\,,\qquad
m_2\,=\,\log\Bigl(\frac{\ga}2\Bigr)\,n_1\,+\,n_2\,,
}\\[0.5cm]
\displaystyle{
m_3\,=\,\log\Bigl(\frac{\ga}2\Bigr)\,n_2\,+\,n_3\,,\qquad
m_4\,=\,\frac{{\cN}^2}{\kappa^2}\,\frac{q_1}4
\,+\,\log\Bigl(\frac{\ga}2\Bigr)\,n_3\,,
}\\[0.5cm]
\displaystyle{ m_5\,=\,n_4\,,\qquad \qquad \qquad \qquad
m_6\,=\,{\cN}^2\,\frac{q_2}4\,+\,
\log\Bigl(\frac{\ga}2\Bigr)\,n_4\,+\,n_5\,. }
\end{array}
\eqno({\rm B}4)
$$
The coefficients $n_6$ and $n_7$, and $m_7$, $m_8$, and $m_9$ are
present in (B1) and (B2) only formally: their explicit values are
not of importance for the present investigation.

Using ({\rm B}1)--({\rm B}4), we estimate the integrals that
constitute $I_{{\cF}}(\rho)$ and obtain
$$
\begin{array}{rcl}
I_{{\cF}}(\rho)\,\simeq\,f_0\,+\,f_1\,\log^2\bigl({\cN}\rho\bigr)
&+&f_2\,\log\bigl({\cN}\rho\bigr)
\\[0.5cm]
+\,\displaystyle{ \rho^2\,\sum_{i=0}^3
f_{7-i}\,\log^i\bigl({\cN}\rho\bigr)} &+&\displaystyle{\rho^4\,
f_9\,\log^3\bigl({\cN}\rho\bigr)\,+\dots } \,,
\end{array}
\eqno({\rm B}5)
$$
where
$$
\begin{array}{l}
\displaystyle{
f_0\,=\,\log\Bigl(\frac{\ga}2\Bigr)\,{\cI}_J\,-\,{\cI}_Y\,,\quad
\qquad f_1\,=\,\frac{q_1}{2 \kappa^2}\,,\quad \qquad
f_2\,=\,{\cI}_J\,,\quad \qquad f_4\,=\,\frac{q_2}{4}\,,
}\\[0.5cm]
\displaystyle{ f_5\,=\,-\frac{{\cN}^2}{\kappa^2}\,\frac{q_1}8\,+\,
\log\Bigl(\frac{\ga}2\Bigr) \frac{n_2}2\,+\,
\Bigl(1\,-\,\log\frac\ga{2}\Bigr) {\h {\cY}}_1\,-\,{\h {\cJ}}_1
\,,
}\\[0.5cm]
\displaystyle{ f_6\,=\,\frac{{\cN}^2}4\,
\left(-\,{\cI}_J\,+\,\frac{q_1}{\kappa^2} \right) \,+\,
\log\Bigl(\frac{\ga}2\Bigr)\,{\h {\cY}}_1\,+\, {\h
{\cY}}_2\,-\,{\h {\cJ}}_2\,,
}\\[0.5cm]
\displaystyle{ f_7\,=\,\frac{{\cN}^2}4\,
\left({\cI}_Y\,+\,\Bigl(1\,-\,\log\frac{\ga}2\Bigr){\cI}_J \right)
\,+\,\log\Bigl(\frac{\ga}2\Bigr)\,{\h {\cY}}_2\,-\,{\h {\cJ}}_3}
\,,\qquad \qquad \displaystyle{f_9\,=\,\frac{n_4}{16}} \,.
\end{array}
\eqno({\rm B}6)
$$
The following notation is accepted in ({\rm B}6):
$$
\begin{array}{l}
\displaystyle{ {\h {\cY}}_1\,=\,\frac{3 q_2}4\,+\,
\frac{n_2\,-\,n_1}2 \,,\qquad {\h {\cY}}_2\,=\,-\frac{{\h
{\cY}}_1}2\,+\, \frac{n_3}2  \,, }
\end{array}
\eqno({\rm B}7)
$$
$$
\begin{array}{l}
\displaystyle{ {\h {\cJ}}_1\,=\,\frac{3 q_2}2\,-\,\frac{3
m_1}4\,+\, \frac{m_2}2\,,\qquad  {\h {\cJ}}_2\,=\,-\,{\h {\cJ}}_1
\,+\,\frac{m_3}2\,,\qquad {\h {\cJ}}_3\,=\,-\,\frac{{\h {\cJ}}_2}2
\,+\,\frac{m_4}2\,. }
\end{array}
\eqno({\rm B}8)
$$
In addition, we denote
$$
\begin{array}{l}
\displaystyle{ {\cI}_J\,\equiv\,{\w {\cI}_J}\,-\,
     \frac{q_1}{\kappa^2}\log{\cN}}\,,\\[0.5cm]
\displaystyle{ {\cI}_Y\,\equiv\,{\w
{\cI}_Y}\,-\,\frac{q_1}{\kappa^2}\log{\cN}\, \Bigl(\frac{\log
{\cN}}2\,+\,\log\frac{\ga}2\Bigr)
     }\,,
\end{array}
\eqno({\rm B}9)
$$
where
$$
\begin{array}{l}
\displaystyle{ {\w {\cI}_J}\,\equiv\,\lim_{\epsilon\to 0+}
\biggl(- \,\int\limits_\epsilon^\infty J_0 ({\cN}t)\,G(\kappa
t)\,t\,dt \,+\,\frac{q_1}{\kappa^2}\int\limits_\epsilon^1
\frac{dt}{t}\biggr)\,,}\\[0.5cm]
\displaystyle{ {\w {\cI}_Y}\,\equiv\,\lim_{\epsilon\to 0+}
\biggl(-\,\int\limits_\epsilon^\infty {\w Y}_0 ({\cN}t)\,G(\kappa
t)\,t\,dt \,+\,\frac{q_1}{\kappa^2}\int\limits_\epsilon^1
\log\Bigl(\frac{\ga}2\,{\cN}t\Bigr)\frac{dt}{t} \biggr)\,.}
\end{array}
\eqno({\rm B}10)
$$
It should be pointed out that possible contributions $\propto
\log^4\bigl({\cN}\rho\bigr)$ are cancelled in (B5): for  instance,
the coincidence of the coefficients $n_4$ and $m_5$ (see (B4))
results in the absence of the corresponding term of the fourth
degree in $\rho$. Moreover, $n_5$ (B3) and $m_6$ (B4) are
necessary to calculate $f_9$ (B6).

Eventually, the estimate (4.7) appears after re-arrangement of
(B5). Generally, the corresponding coefficients $r_0, r_1, \dots,
r_7$ look as follows:
$$
\begin{array}{l}
r_0\,=\,f_0\,+\,f_1\,\log^2{\cN}\,+\,f_2\,\log{\cN}\,,\quad\qquad
r_1\,=\,2\,f_1\,\log{\cN}\,+\,f_2 \,,
\\[0.5cm]
r_2\,=\,f_1\,,\qquad\qquad r_3\,=\,f_4\,, \qquad\qquad\qquad
r_4\,=\,3\,f_4\,\log{\cN}\,+\,f_5\,,
\\[0.5cm]
r_5\,=\,3\,f_4\,\log^2{\cN}\,+\, 2\,f_5\,\log{\cN}\,+\,f_6\,,
\\[0.5cm]
\displaystyle{ r_6\,=\,f_4\log^3{\cN}\,+\,f_5
\log^2{\cN}\,+\,f_6\,\log {\cN}\,+\,f_7}\,,
\\[0.5cm]
\displaystyle{ r_7 \,=\,f_9\,=\, ({\kappa}^2\,-\,{\cN}^2)
\frac{p_2}{384}\,. }
\end{array}
\eqno({\rm B}11)
$$
We obtain $r_0, r_1,r_2$, and $r_3$ by means of (B6).  To obtain
$r_4$ (4.8), we re-express $f_5$:
$$
f_5\,=\, \Bigl(1\,-\,\frac{{\cN}^2}{\kappa^2}\Bigr)\,
\frac{p_1}{32}\,-\, \Bigl(1\,+\,\log\frac{{\cN}}{\kappa}\Bigr)
\,\frac{p_2}8\,+\,\frac{p_3}8 \,. \eqno({\rm B}12)
$$

Now we turn to $r_5$ and $r_6$ from (5.10). Although $f_6$ and
$f_7$ are left undone, their dependence on $const$ can be
extracted by means of (B6)--(B10) as follows:
$$
f_6\,=\,- \kappa^2\,\frac{{\w C}}{4 k}\,+\,\dots ,\qquad
f_7\,=\,\kappa^2 (q\,+\,\log {\cN})\frac{{\w C}}{4 k}\,+
\,\dots\,, \eqno({\rm B}13)
$$
where ${\w C}$ and $q$ are determined by (5.11). Since $f_4$ (B6)
and $f_5$ (B12) are free from $const$, the equation (5.10) is
indeed valid.

\end{document}